\documentclass[11pt,onecolumn,letter]{article}

\usepackage{amssymb}
\usepackage{graphicx}
\usepackage{dcolumn}
\usepackage{bm}
\usepackage[cdot,mediumqspace,amssymb]{SIunits} 
\usepackage[autolanguage]{numprint} 
\usepackage{ifthen}
\usepackage{ifpdf}
\usepackage{rotating} 
\usepackage[ruled,vlined,linesnumbered]{algorithm2e} 
\usepackage{rotating}
\usepackage{anysize}
\usepackage{xcolor} 
\usepackage{tabularx}
\usepackage{amsmath}

\ifpdf
    \graphicspath{{Figures/PDF/}}
\else
    \graphicspath{{Figures/EPS/}}
\fi

\usepackage{cleveref}
\marginsize{0.75in}{0.75in}{0.75in}{0.75in}

\begin{document}

\title{ Resolving structural variability in network models and the brain }

\author{Florian Klimm$^{1,2,3}$, Danielle S. Bassett$^{1,4,*}$, Jean M. Carlson$^{1}$, Peter J. Mucha$^{3,5}$\\
\small $^{1}$Department of Physics, University of California, Santa Barbara, CA 93106 USA\\
\small $^{2}$ Institut f\"{u}r Physik, Humboldt-Universit\"{a}t zu Berlin, 12489 Germany\\
\small $^{3}$ Department of Mathematics, University of North Carolina, Chapel Hill, NC 27599 USA\\
\small $^{4}$ Sage Center for the Study of the Mind, University of California, Santa Barbara, CA 93106 USA\\
\small $^{5}$ Department of Applied Physical Sciences, University of North Carolina, Chapel Hill, NC 27599 USA\\
\small $^{*}$Corresponding author email dbassett@physics.ucsb.edu}
\maketitle

\normalsize

\begin{abstract}
Large-scale white matter pathways crisscrossing the cortex create a complex pattern of connectivity that underlies human cognitive function. Generative mechanisms for this architecture have been difficult to identify in part because little is known in general about mechanistic drivers of structured networks. Here we contrast network properties derived from diffusion spectrum imaging data of the human brain with 13 synthetic network models chosen to probe the roles of physical network embedding and temporal network growth. We characterize both the empirical and synthetic networks using familiar graph metrics, but presented here in a more complete statistical form, as scatter plots and distributions, to reveal the full range of variability of each measure across scales in the network. We focus specifically on the degree distribution, degree assortativity, hierarchy, topological Rentian scaling, and topological fractal scaling---in addition to several summary statistics, including the mean clustering coefficient, the shortest path-length, and the network diameter. The models are investigated in a progressive, branching sequence, aimed at capturing different elements thought to be important in the brain, and range from simple random and regular networks, to models that incorporate specific growth rules and constraints. We find that synthetic models that constrain the network nodes to be physically embedded in anatomical brain regions tend to produce distributions that are most similar to the corresponding measurements for the brain. We also find that network models hardcoded to display one network property (e.g., assortativity) do not in general simultaneously display a second (e.g., hierarchy). This relative independence of network properties suggests that multiple neurobiological mechanisms might be at play in the development of human brain network architecture. Together, the network models that we develop and employ provide a potentially useful starting point for the statistical inference of brain network structure from neuroimaging data.
\end{abstract}

\newpage
\section{Introduction}
\addcontentsline{toc}{section}{Introduction}

Increasing resolution of noninvasive neuroimaging methods for quantifying structural brain organization in humans has inspired a great deal of theoretical activity \cite{Sporns2013,Kaiser2011,Alstott2009,Honey2009}, aimed at developing methods to understand, diagnose, and predict aspects of human development and behavior based on underlying organizational principles deduced from these measurements \cite{Griffa2013,Fornito2012,Behrens2012}. Ultimately, the brain is a network, composed of neuronal cell bodies residing in cortical grey matter regions, joined by axons, protected by myelin. Diffusion-weighted magnetic resonance imaging methods trace these white matter connections, based on the diffusion of water molecules through the axonal fiber bundles. While resolution has not reached the level of individual neurons and axons, these methods lead to reliable estimates of the density of connections between regions and fiber path lengths. The result is a weighted adjacency matrix, with a size and complexity that increases with the resolution of the measurements \cite{Bassett2010c,Cammoun2012}.

The immense complexity of this data makes it difficult to directly deduce the underlying mechanisms that may lead to fundamental patterns of organization and development in the brain \cite{Hermundstad2013}. As a result, comparison studies with synthetic network models, employing quantitative graph statistics to reduce the data to a smaller number of diagnostics, have provided valuable insights \cite{Rubinov2011b,Bassett2009,Supekar2008,He2007,Achard2006}. These models and statistics provide a vehicle to compare neuroimaging data with corresponding measurements for well-characterized network null models. However, the methods are still in development \cite{vanWijk2010,Fornito2013,Johansenberg2013}, and vulnerable to the loss of critical information through oversimplification of complex, structured data sets, by restricting comparisons to coarse measurements that ignore variability \cite{Simpson2012,Rubinov2011c,Hermundstad2013}.

Two critical questions motivate development of network methodologies for the brain. The first question focuses on predictive statistics: Are there graph metrics that may ultimately be useful in parsing individual differences and diagnosing diseases?  Comparing empirical brain data to benchmark null models might help establish statistical significance of a topological property \cite{Humphries2006,Bullmore2011,Bassett2012c} or assist in obtaining a statistical inference about differences in brain network structure between groups \cite{vanWijk2010}. The second question focuses on network characteristics from a fundamental, development and evolutionary perspective: What organizational principles underlie growth in the human brain? Here comparing empirical brain data to simplified model networks that have been created to capture some aspect of, for example, neurodevelopmental growth rules \cite{Cahalane2011}, neuronal functions \cite{Rubinov2011b}, or physiological constraints \cite{Vertes2012} may aid in developing a mechanistic understanding of the brain's network architecture (e.g., \cite{Kaiser2006,Chen2006,Bassett2010}). Both efforts require a basic understanding of the topological similarities and differences between synthetic networks and empirical data.

\begin{figure}[h!]
\begin{center}
\includegraphics[width=0.98\textwidth]{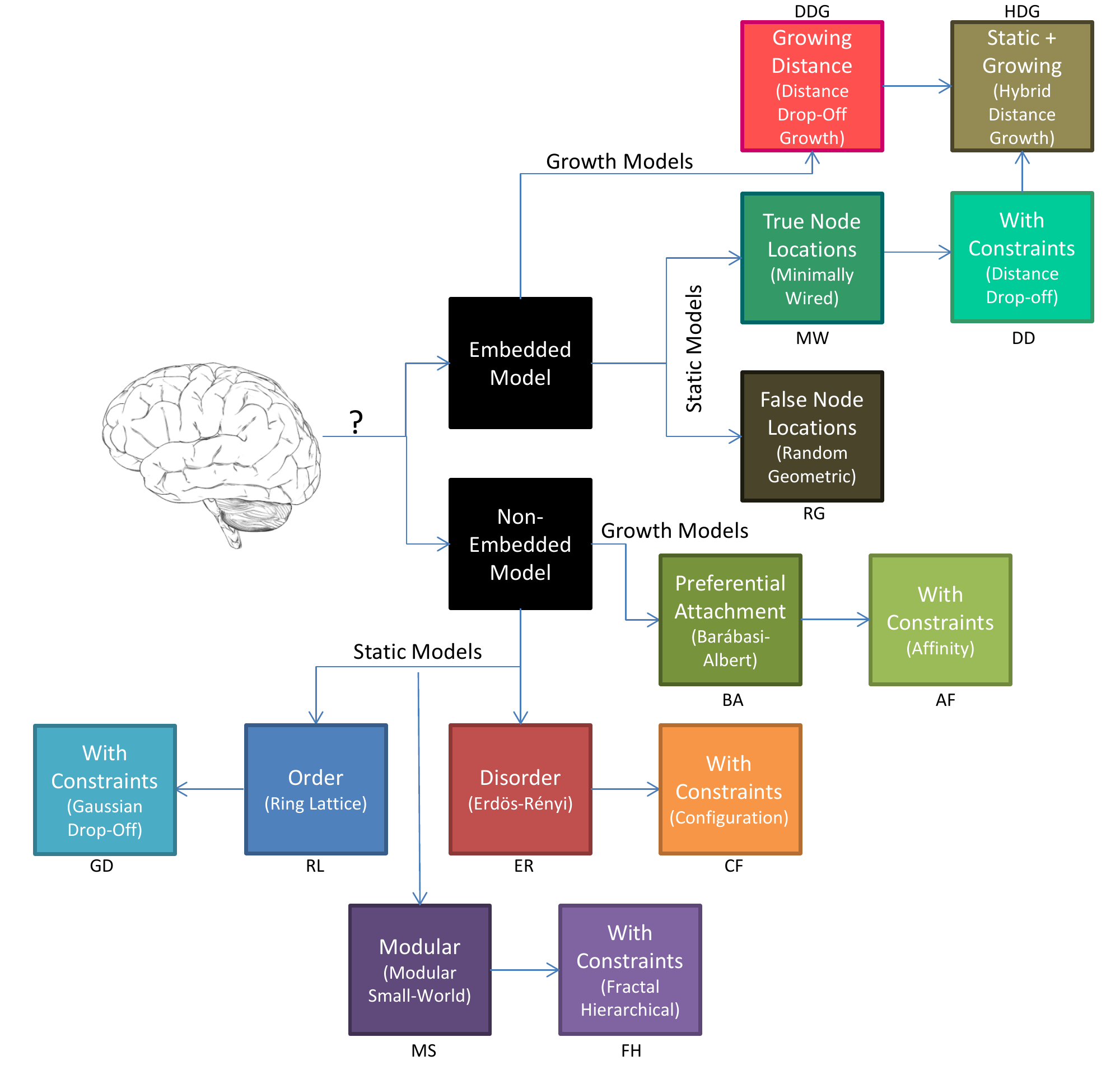}
 \caption{ \textbf{Branching Structure of Synthetic Model Examination.} We distinguish between synthetic networks that are constructed based on rules for connectivity between nodes (\emph{non-embedded}), and those that constrain nodes to reside in anatomical brain regions (\emph{embedded}). We further distinguish between synthetic networks that are obtained from static ensembles (\emph{static}), and those that are obtained from growth rules (\emph{growing}). In the non-embedded case, we explore common benchmark networks including regular lattice, Erd\"{o}s-R\'{e}nyi, and small-world models as well as a second set of networks that are based on these benchmarks but that also employ additional constraints. For growing models, we explore the Bar\'{a}basi-Albert model and introduce an \emph{affinity} model inspired by preferential attachment-like properties of neuronal growth. In the embedded case, we distinguish between models that utilize true or false node locations (i.e., models derived from a spatial embedding independent of the known, physical node locations) and explore several growing models inspired by hypotheses regarding wiring minimization in brain development \cite{Kaiser2006,Bassett2010,Bullmore2012}.} \label{Fig1}
\end{center}
\end{figure}

In this paper, we perform a sequence of detailed, topological comparisons between empirical brain networks obtained from diffusion imaging data and 13 synthetic network models. The models are investigated in a tree-like branching order, beginning with the simplest, random or regular graphs, and progressively adding complexity and constraints (see Figure \ref{Fig1}). The objective of this investigation is to determine, in a controlled, synthetic setting, the impact of additional network properties on the topological measurements.

At the coarsest level in the model hierarchy, we distinguish between synthetic networks that are constructed purely based on rules for connectivity between nodes (non-embedded), and those that constrain nodes to reside in anatomical brain regions (embedded) (see Figure \ref{Fig1}). While non-embedded models are frequently used for statistical inference, recent evidence has suggested that physical, embedding constraints may have important implications for the topology of the brain's large-scale anatomical connectivity \cite{Kaiser2006,Chen2006,Kaiser2011,Bassett2010,Bassett2010c,Bullmore2011,Bullmore2012}. By examining both non-embedded and embedded models, we hope our results will help to guide the use, development, and understanding of more biologically realistic models for both statistical and mechanistic purposes \cite{Bassett2011a,Bassett2012c}.

A second important classification of the synthetic models in our study separates those obtained from static ensembles with fixed statistical properties and those generated using mechanistic growth rules (see Figure \ref{Fig1}). While algorithms for generating networks based on static sampling and growth rules ultimately both produce ensembles of fixed graphs for our comparison with data, additional constraints imposed by underlying growth rules may facilitate understanding of mechanisms for development and evolution in the brain as well as other biological and technological networks.

To compare the models with brain data, we employ a subset of the many network diagnostics that have been proposed as measures of network topology \cite{Newman2010}. Many network diagnostics can be described as \emph{summary diagnostics}, in which a property of the network organization is reduced to a single diagnostic number. However, the comparison of summary diagnostics between real and model networks can be difficult to interpret \cite{AlexanderBloch2012} because they often hide the granularity at which biological interpretations can be made. To maximize the potential for a mechanistic understanding, we therefore study the following four diagnostic relationships obtained from a distribution of values over network nodes or topological scales: hierarchy \cite{Ravasz2003}, degree assortativity \cite{Newman2006}, topological Rentian scaling \cite{Christie2000,Greenfield2010}, and the topological fractal dimension \cite{Song2005}. Each of these relational properties have previously been investigated in the context of anatomical brain networks in humans \cite{Bassett2008,Hagmann2008,Bassett2010}. In this paper, we use them to examine the differences between empirically derived anatomical brain networks and synthetic network models.

\begin{table}[h]
\begin{center}
\begin{tabularx}{\textwidth}{| l | l | X | l |}
\hline
Model Name            & Abbreviation  & Description & Citation \\
\hline
\hline
\textbf{Non-embedded}    &  ~            & ~           & ~        \\
\hline
\hline
\emph{Static}   &  ~            & ~           & ~        \\
\hline
Erd\H{o}s-R\'{e}nyi & ER        & Uniform connection probability & \cite{Bollobas2001} \\
Configuration   & CF            & Random rewiring preserving degree distribution & \cite{Newman2003}\\
Ring Lattice    & RL            & Fixed degree to $k$ nearest neighbors & \cite{Bollobas2001} \\
Gaussian Drop-Off & GD          & Gaussian drop-off in edge density with increasing distance from the diagonal & \cite{May2011,Rubinov2009} \\
Modular Small-World & MS        & Fully connected modules linked together by evenly distributed random connections & \cite{Rubinov2009} \\
Fractal Hierarchical & FH       & Modular structure across $n$ hierarchical levels; connection density decays as $1/(E^n)$ & \cite{Rubinov2009} \\
\hline
\emph{Growth}   &  ~            & ~           & ~        \\
\hline
Barab\'{a}si-Albert & BA        & Network growth by preferential attachment rule & \cite{BA2002} \\
Affinity            & AF        & Two-step preferential attachment growth with hardcoded assortativity and hierarchy & ~ \\
\hline
\textbf{Embedded}    &  ~            & ~           & ~        \\
\hline
\hline
\emph{Static}   &  ~            & ~           & ~        \\
\hline
Random Geometric & RG           & Wire together random node locations with shortest possible connections & \cite{Barthelemy2010} \\
Minimally Wired  & MW           & Wire together true node locations with shortest possible connections & \cite{Bassett2010,Kaiser2006,Chen2006} \\
Distance Drop-Off& DD           & Wire together true node locations with a probability that drops off with distance between nodes & \cite{Avin2008} \\
\hline
\emph{Growth}   &  ~            & ~           & ~        \\
\hline
Distance Drop-Off Growth & DDG   & Network growth by distance drop-off rule & ~ \\
Hybrid Distance Growth & HDG      & Minimally wired network that grows with distance drop-off rule & ~ \\
\hline
\end{tabularx}
\end{center}
\vspace{-5mm} \caption{\textbf{Network Models Names, Abbreviations, Intuitive Descriptions, and Associated References.} \label{Tab1}}
\end{table}

\newpage
\section{Materials and Methods} \label{sec:methods}
\addcontentsline{toc}{section}{Materials and Methods}
\subsection{Data}
\begin{figure}[h!]
\begin{center}
\includegraphics[width=0.98\textwidth]{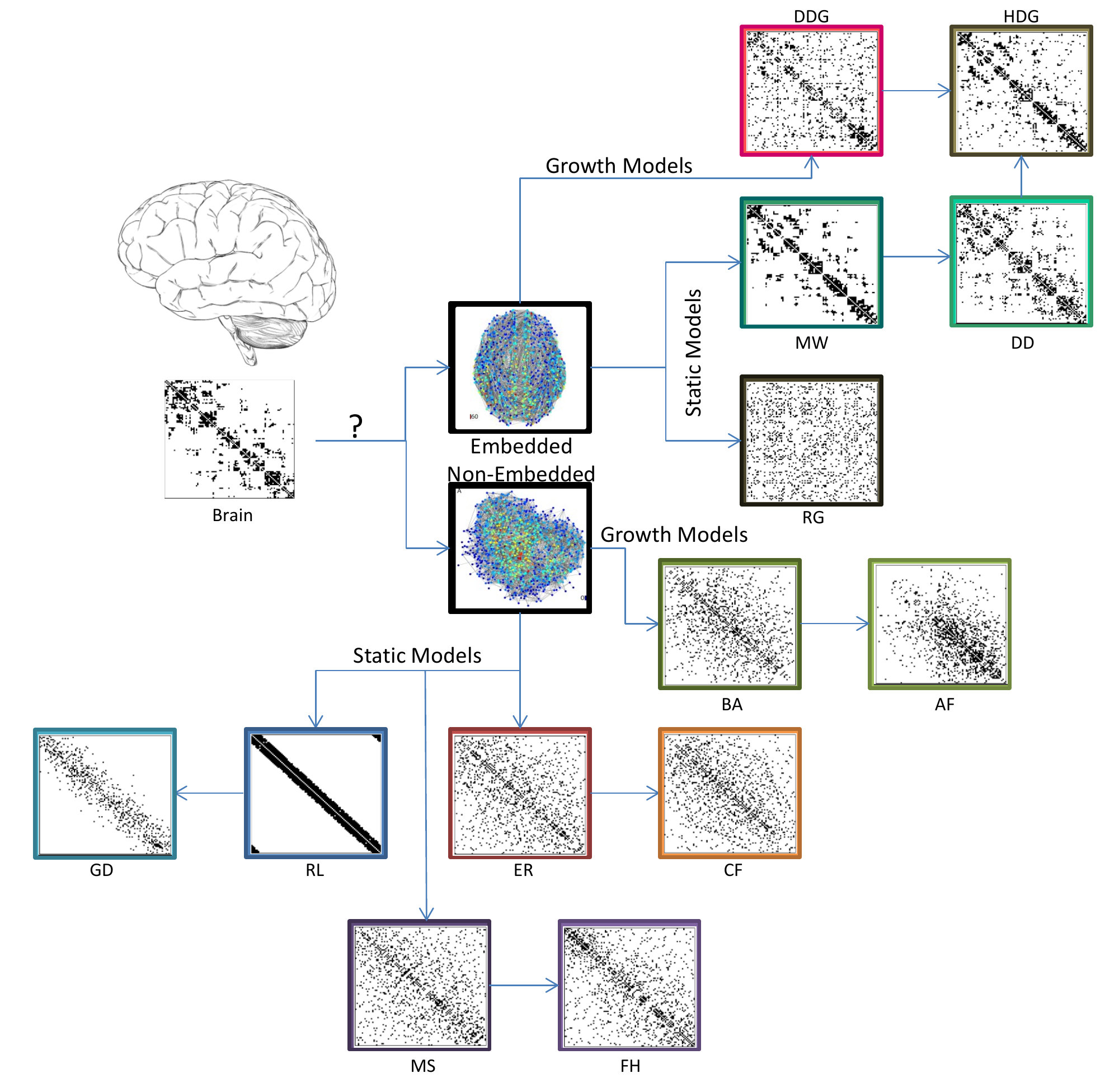}
 \caption{ \textbf{Adjacency Matrices for Brain and Synthetic Models.} Example adjacency matrices are provided for the brain and for the 13 synthetical network models described in Figure \ref{Fig1}. In the empirical brain data and the non-embedded null models, network nodes are ordered along the $x$ and $y$-axes to maximize connectivity along the diagonal, as implemented by the \textit{reorderMAT.m} function in the Brain Connectivity Toolbox \cite{Rubinov2009}. In the embedded models, nodes are listed in the same order as they are in the empirical brain data. Abbreviations are as listed in Table \ref{Tab1}.} \label{Fig2}
\end{center}
\end{figure}

We utilize previously published diffusion spectrum imaging data \cite{Hagmann2008} to examine network structure of anatomical connectivity in the human brain. The direct pathways between $N=998$ regions of interest are estimated using deterministic white matter tractography in 5 healthy human participants \cite{Hagmann2008}. This procedure results in an $N\times N$ weighted undirected adjacency matrix $\mathbf{W}$ representing the network, with elements $W_{ij}$ indicating the (normalized) number of streamlines connecting region $i$ to region $j$ (see Figure \ref{Fig2}). The organization of white matter tracts can be examined at two distinct levels of detail: topological and weighted.
Studies of the topological organization of brain anatomy focus on understanding the presence or absence of white matter tracts between regions \cite{Chen2006,Kaiser2006,Bassett2010}, while studies of the weighted organization focus on understanding the strength of white matter connectivity between those regions. In this paper, we explore the topological organization of white matter connectivity. In future work we plan to build additional constraints into our models that will enable a comparison of model and empirical weighted networks.

To study topological organization, we construct the binary adjacency matrix $\mathbf{A}$ in which the element $A_{ij}$ is equal to $1$ if the employed tractography algorithm identifies any tracts (of any strength) linking region $i$ with region $j$ (i.e., $W_{ij} \neq 0$).  In this data \cite{Hagmann2008}, the adjacency matrix $\mathbf{A}$ is relatively sparse, resulting in a network \textit{density} of $\rho=2M/[N(N-1)]\approx 2.7\%$, where $M=\frac{1}{2}\sum_{ij}A_{ij}$ is the total number of connections present. This estimate of brain network sparsity is consistent with estimates extracted from other similar data sets of comparable network size \cite{Bassett2010c,Zalesky2010}.

Given the potential variability in the topological organization of networks extracted from different individuals \cite{Bassett2010c,Dennis2012,Owen2013,Cheng2012,Vaessen2010}, we report results for one individual in the main manuscript and describe the consistency of these results across subjects in the Supplementary Materials.

\subsection{Network Diagnostics}
\addcontentsline{toc}{subsection}{Network Diagnostics}

We measure four network properties including degree assortativity, hierarchy, Rentian scaling, and topological fractal dimension as well as several summary diagnostics, as reported in Table \ref{tb:metrics}.

\textbf{Assortativity.}
The number of edges emanating from node $i$ is referred to as its degree, denoted by $k_{i}$. The degree assortativity of a network, or more simply `assortativity' here, is defined as the correlation between a node's degree and the mean degrees of that node's neighbors which can be calculated as
\begin{equation}
 r = \frac{M^{-1} \sum_{m} j_{m} k_{m} - [M^{-1} \sum_{m} \frac{1}{2} (j_{m} + k_{m})]^{2} } {M^{-1} \sum_{m} \frac{1}{2} (j_{m}^{2} + k_{m}^{2}) - [M^{-1} \sum_{m} \frac{1}{2} (j_{m}+k_{m})]^{2}},
\end{equation}
where $j_{m},k_{m}$ are the degrees of the nodes at either end of the $m^{th}$ edge, with $m=1\ldots M$ \cite{Newman2002}. The assortativity measures the likelihood that a node connects to other nodes of similar degree (leading to an assortative network, $r>0$) or to other nodes of significantly different degree (leading to a disassortative network, $r<0$). Social networks are commonly found to be assortative while networks such as the internet, World-Wide Web, protein interaction networks, food webs, and the neural network of \textit{C. elegans} are disassortative \cite{Newman2006}.

\textbf{Hierarchy.}
The hierarchy of a network is defined quantitatively by a relationship between the node degree and the local clustering coefficient $C_i$ \cite{Watts1998}.  For each individual node $i$, $C_i$ is defined as:
\begin{equation}
	C_i=\frac{\Delta _{exist}}{\Delta _{possible}}
\end{equation}
where $\Delta_{exist}$ is the ratio of the number of existing triangle subgraphs that include node $i$, and $\Delta_{possible}$ is the number of node triples containing node $i$. Using this local definition, the clustering coefficient of the graph $C$ as a whole (a summary diagnostic) is defined as the mean of $C_i$ over all nodes in the network.

The definition of hierarchy is based on a presumed power law relationship between the local clustering coefficient $C_i$ and the degree $k_i$ of all
nodes $i$ in the network \cite{Ravasz2003}:
\begin{equation}\label{hierarchy}
C_i \sim k_i^{- \beta} .
\end{equation}
For a given network, the best fit to the scaling exponent $\beta$ is referred to as the network hierarchy.

\textbf{Topological Rentian Scaling.}
In contrast to the physical Rent's rule \cite{Christie2000}, the topological Rent's rule is defined as the scaling of the number of nodes $n$ within a topological partition of a network with the number of connections or edges, $e$, crossing the boundary of that topological partition. If the relationship between these two variables is described by a power law (i.e., $e \propto n^{p_{T}}$), the network is said to show topological Rentian scaling, or a fractal topology, and the exponent of this scaling relationship is known as the topological Rent exponent, $p_{T}$ \cite{Ozaktas1992}. Thus higher values of the topological Rent exponent are indicative of higher dimensional network topology. Pragmatically, to determine $p_{T}$, we follow the procedure outlined in \cite{Greenfield2010} where topological partitions are created by a recursive min-cut bi-partitioning algorithm that ignores spatial locations of network nodes \cite{Bassett2010}.

\textbf{Topological Fractal Dimension.}
The topological Rent's exponent described above is related to the topological dimension, $D_{T}$, of the network according to the inequality $p_{T} \geq 1 - \frac{1}{D_{T}}$ \cite{Ozaktas1992}. To directly quantify the topological dimension of a network, we evaluate its topological invariance under length-scale transformations \cite{Song2005}. We employ a box-counting method \cite{Concas2006} in which we count the number of boxes $N_B$ of topological size $l_B$ that are necessary to cover the network. The fractal dimension of the network can then be estimated as the exponent $d_B$ of the putative power law relationship
\begin{equation}
	N_B\approx l_N^{-d_B}.
\end{equation}
The fractal dimension of a network is a measure of the network's complexity. We note that the process of tiling the network into boxes of different sizes is non-deterministic. To account for this variability, we report mean values of $d_{B}$ over 50 different tilings of the a given network.

\textbf{Additional Quantities of Interest.}
In Table \ref{tb:metrics}, we list several summary diagnostics of interest to complement our analysis of relational properties. These include the average path length between node $i$ and $j$, defined as the shortest number of edges one would have to traverse to move from node $i$ to node $j$ \cite{Dijkstra1959}. The path length of an entire network, $P$, is then defined as the average path length from any node to any other node in the network: $P=\frac{1}{N(N-1)} \sum_{ij} P_{ij}$, while the maximal path length between any two pairs of nodes is called the \emph{diameter} $D=\max_{ij}\{P_{ij}\}$.

\subsection{Statistics, Software, and Visualization}
\addcontentsline{toc}{subsection}{Statistics, Software, and Visualization}

All computational and basic statistical operations (such as t-tests and correlations) were implemented using MATLAB$^{\textregistered}$ (2009b, The MathWorks Inc., Natick, MA) software. Graph diagnostics were estimated using a combination of in-house software, the Brain Connectivity Toolbox \cite{Rubinov2009}, and the MATLAB Boost Graph Library (\verb http://www.stanford.edu/~dgleich/programs/ ). To perform the recursive topological partitioning employed in the examination of topological Rentian scaling, we used the software tool \textit{hMETIS} \cite{hmetis}.

Several of the network models that we investigate include one or more tunable parameters affecting the details of the generated graphs. These include the Barab\'{a}si-Albert, affinity, and hybrid distance growth models. To compare these network models to the data, we optimized parameter values to minimize the difference between the model network and the empirical brain network. Specifically, we used the Nelder-Mead simplex method, which is a derivative-free optimization method, that minimizes the value of a difference metric $\delta_{m}$ between the two networks. We chose to let $\delta_{m}$ be the sum of the absolute relative difference of seven of the network characteristics reported in Table \ref{tb:metrics} (clustering coefficient $C$, path length $P$, diameter $D$, degree assortativity $r$, hierarchical parameter $\beta$, topological Rentian exponent $p_{T}$, and topological fractal dimension $d_{B}$), although we note that other choices are of course possible.

\clearpage
\newpage
\section{Results}
\addcontentsline{toc}{section}{Results}

In this section we individually compare topological network diagnostics calculated for the empirical brain data to each of the 13 network models that appear in Figures~\ref{Fig1} and \ref{Fig2}. We proceed through the catalog of synthetic models along the branches illustrated in Figure \ref{Fig1}. We begin with the simplest models (i.e. non-embedded, static, random and regular), and incrementally add structure, constraints, growth mechanisms, and embedding in order to isolate how these additional features impact the measured diagnostics.

\begin{figure}[h!]
\begin{center}
\includegraphics[width=0.66\textwidth]{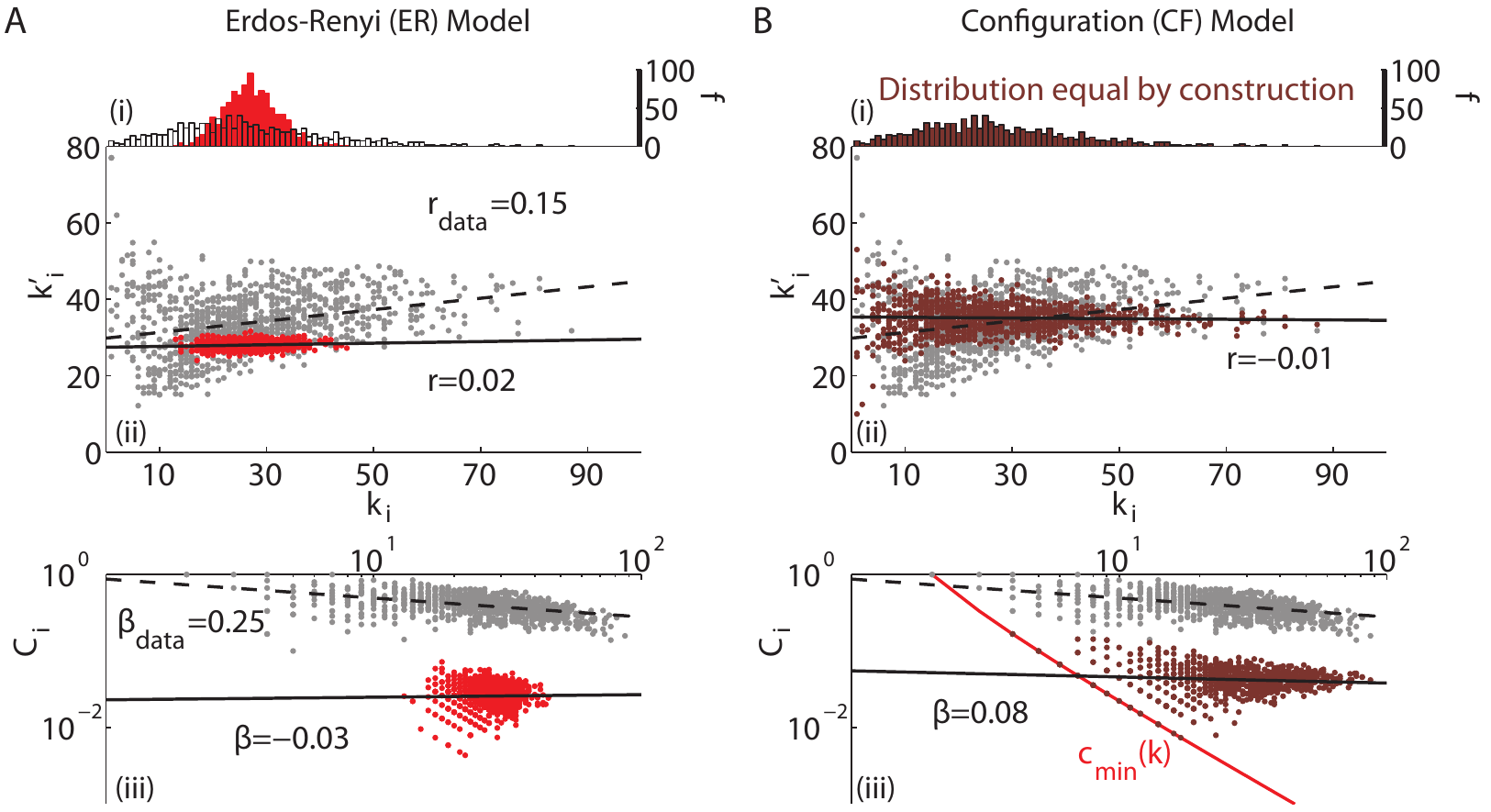}
\caption{Comparison between the \emph{(i)} degree distribution (number $f$ of nodes with a given degree $k_{i}$), \emph{(ii)} assortativity (correlation between a node's degree $k_{i}$ and the mean degree of that node's neighbors $k'_{i}$, summarized by parameter $r$), and \emph{(iii)} hierarchy (the relationship between the clustering coefficient $C_{i}$ and the degree $k_{i}$ over all nodes in the network, summarized by parameter $\beta$) of the \emph{(A)} Erd\H{o}s-R\'{e}nyi and \emph{(B)} configuration model with conserved degree distribution models and the same diagnostics of the brain anatomical data (grey). Black lines indicate best linear fit to the data (dashed) and model (solid) networks. In panel \emph{(B)} the lower (nonzero) bound on the clustering coefficient---which corresponds to the presence of only one triangle---as a function of degree is indicated by the red line.}
\label{fig:ER}
\end{center}
\end{figure}

For each network we present statistical results for three diagnostics: (i) the degree distribution $P(k_i)$ vs. $k_i$, (ii) the mean node degree of the neighboring nodes vs.~node degree $k_i$ for each node $i$ (used to calculate assortativity), and (iii) the local clustering coefficient $C_i$ vs.~node degree $k_i$ for each $i$ (used to calculate hierarchy). In Figures \ref{fig:ER}--\ref{fig:BA} and \ref{fig:RGG}--\ref{fig:GRDG}, the results for the empirical brain network are shown in gray and the corresponding results for each of the synthetic models are shown in a contrasting color on the same graph to facilitate comparisons. In addition, we illustrate the scaling relationships used to evaluate Rentian scaling and the topological dimension of each network (see Figures \ref{fig:fracdim} and \ref{fig:fracdimphys}).

For our comparisons, we group the models first into the set of non-embedded models, followed by the embedded models and we further group results according to the branches of inquiry outlined in Figures~\ref{Fig1} and ~\ref{Fig2}. For each model we briefly describe our method for generating the synthetic network, followed by a description of the diagnostics compared to the empirical results.

\subsection{Non-Embedded Network Models}
We begin by comparing the network organization of the brain's anatomical connectivity with that of 8 network models whose structure is not {\it a priori} constrained to accommodate a physical embedding of the nodes onto cortical regions in the brain. (In the next subsection, we will examine 5 embedded network models.) The non-embedded network models include an Erd\H{o}s-R\'{e}nyi graph, a configuration model with the same degree distribution as the empirical network, a ring lattice graph, a modular small-world graph, a fractal hierarchical graph, a Gaussian drop-off model, a Barab\'asi-Albert graph, and an affinity model (see Figure \ref{Fig2} for associated example adjacency matrices for these graphs and Table \ref{Tab1} for abbreviations of model names). These models range from disordered to ordered (e.g., the Erd\H{o}s-R\'{e}nyi and regular lattice models) with a range of mesoscale organization for intermediate cases (e.g., modular small-world and fractal hierarchical models) which influence the network diagnostics, and (dis)similarities to corresponding measurements for the brain.

\subsubsection{Static Non-Embedded Models}

\textbf{Erd\H{o}s-R\'{e}nyi (ER) Model:} The Erd\H{o}s-R\'{e}nyi (ER) model is an important benchmark network that is often used as a comparison null mode for statistical inference. Specifically, we consider the `$G(N,M)$ model' where the ER graph is constructed by connecting pairs chosen uniformly at random from $N$ total nodes until $M$ edges exist in the graph \cite{Bollobas2001}. The degree distribution generated by this procedure is, as expected, relatively symmetric about the mean degree $\rho(N-1) \approx 27$ (see Figure \ref{fig:ER}A(i)).

The ER model is a poor fit to brain anatomical connectivity (see Figure \ref{fig:ER}A). The degree distribution is much more sharply peaked than the corresponding distribution for the brain. For the ER graph the variance is approximately equal to the mean degree, while the corresponding data for the brain is more broadly distributed. As a result, the ER network misses structure associated with both high degree hubs and low degree nodes. Because edges are placed at random, organizational properties like assortativity and hierarchy are not observed and---as expected theoretically---the clustering coefficient is smaller and the path length shorter than that of anatomical brain networks (see Table \ref{tb:metrics}).

\textbf{Configuration (CF) Model}: We next consider a modification of the ER graph that is constrained to have the same degree distribution as the empirical data. We refer to this as the configuration model (CF). We generate randomized graphs by an algorithm that chooses two existing connections uniformly at random ($a \longleftrightarrow b$ and $c \longleftrightarrow d$) and switches their associations ($a \longleftrightarrow d$ and $c \longleftrightarrow b$) \cite{Maslov2002}.

This model agrees with the empirical degree distribution by construction (see Figure \ref{fig:ER}B(i)). However, it does not fit the higher order association of a node's degree with that node's mean neighbor degree (assortativity) (see Figure \ref{fig:ER}B(ii)). The average clustering coefficient remains small, although it is larger than that observed in the ER network. In Figure \ref{fig:ER}B(iii), we observe a small association between the clustering coefficient and degree (hierarchy) which appears to be driven by nodes of small degree. To interpret this finding, we note that the nonzero minimum of the clustering coefficient of a node of degree $k$ is given by
\begin{equation}
c_{\mathrm{min}\neq0}(k)=\frac{2}{k(k-1)}.
\end{equation}
Thus, nodes of small degree tend to have a higher minimum non-zero clustering than nodes of high degree. In comparison to the ER model, the existence of small degree nodes leads to an increased diameter of the graph whereas the existence of high degree nodes leads to the maintenance of a short average path length.

\begin{figure}[h!]
\begin{center}
\includegraphics[width=0.66\textwidth]{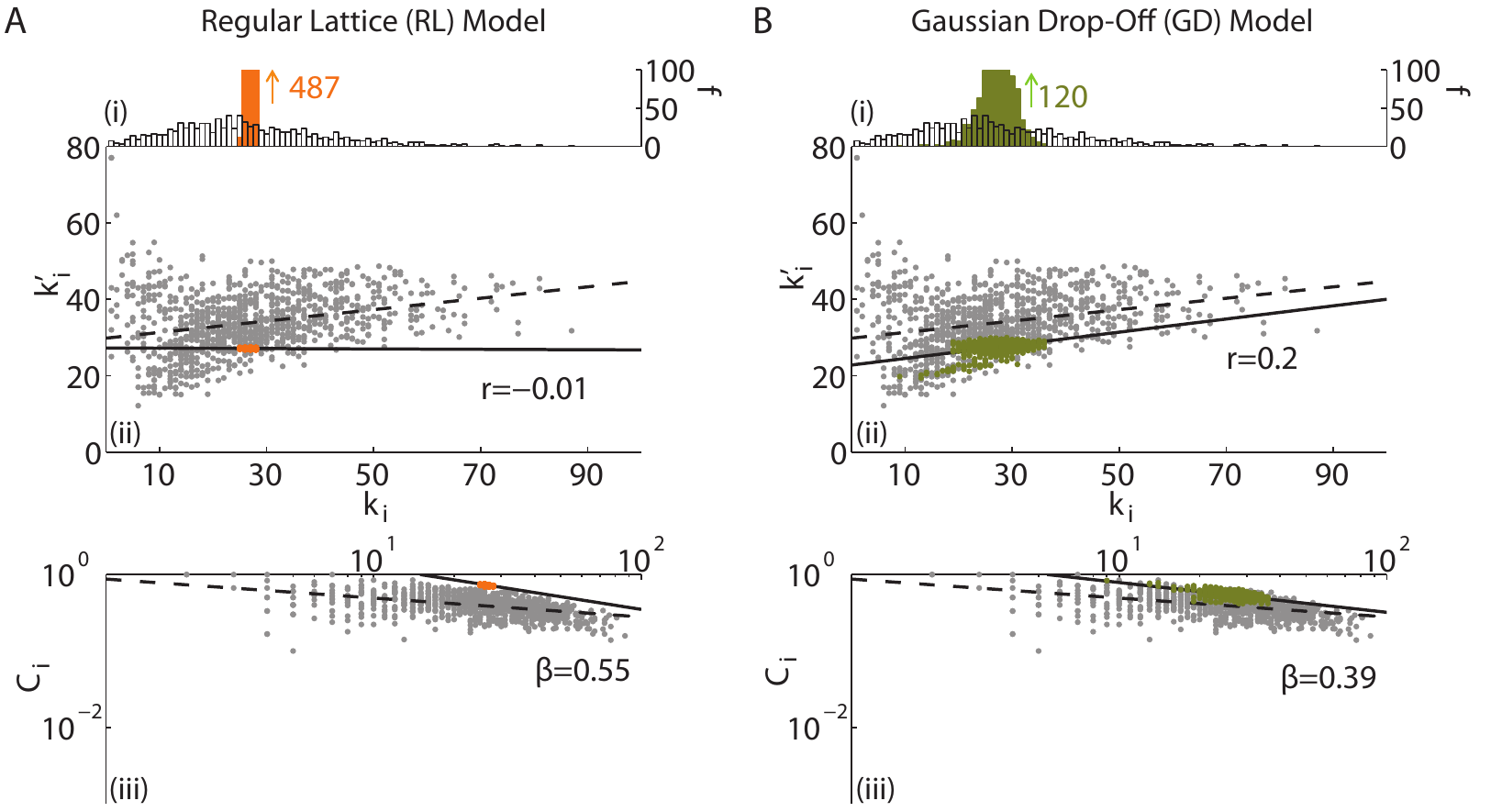}
\caption{Comparison between the \emph{(i)} degree distribution (number $f$ of nodes with a given degree $k_{i}$), \emph{(ii)} assortativity (correlation between a node's degree $k_{i}$ and the mean degree of that node's neighbors $k'_{i}$, summarized by parameter $r$), and \emph{(iii)} hierarchy (the relationship between the clustering coefficient $C_{i}$ and the degree $k_{i}$ over all nodes in the network, summarized by parameter $\beta$) of the \emph{(A)} ring lattice and \emph{(B)} Gaussian drop-off models and the same diagnostics in the brain anatomical data (grey). Black lines indicate best linear fit to the data (dashed) and model (solid) networks.}
\label{fig:RL}
\end{center}
\end{figure}

\textbf{Ring Lattice (RL) Model:} In contrast to the two previous models, the ring lattice (RL) model has a highly ordered topology where each node is connected to its $\frac{2M}{N}\approx 27$ nearest neighbors.

By construction, the degree distribution for the ring lattice is extremely sharply peaked. If the number of edges $M$ is divisible by the number of nodes $N$, then all nodes have equal degree, otherwise the remainder is distributed uniformly at random throughout the network, resulting is a very narrow spread in the distribution. The clustering coefficient is close to unity, indicating that most neighbors of a node are also connected to each other. The restriction to local connectivity results in a large diameter and long average path length. The small variation in degree induced by the random distribution of the remaining edges is insufficient to induce assortativity (see Figure \ref{fig:RL}A). Interestingly, however, this model displays topological network hierarchy because nodes that have been assigned those remaining edges have a higher than average degree which directly decreases the clustering coefficient of those nodes. These results underscore the fact that very small amounts of noise in a data set can create the illusion of the presence of a network property where it does not exist.

\textbf{Gaussian Drop-Off (GD) Model:}
Compared to the brain, the random and randomized models exhibit lower clustering, and the regular ring lattice exhibits higher clustering. An intermediate topology between these two extremes is obtained by generalizing the concept of local connections from the ring lattice to a stochastically generated network where the density of connections drops off at rate $\kappa$ with increasing distance from the main diagonal of the adjacency matrix.

We chose a value for $\kappa$ by examining the empirical brain data as follows. First, we reordered the adjacency matrix such that the connections (represented by nonzero matrix elements) are predominantly located near the matrix diagonal, using the code {\tt reorderMAT.m} in the Brain Connectivity Toolbox \cite{Rubinov2009}. We then fit a Gaussian function to the empirical drop-off of the first 400 off-diagonal rows of the reordered brain adjacency matrix \cite{Rubinov2009}. We note that the fit provided an $R^2$ value of approximately $0.75$.

The very localized structure in this model, similar to that observed in an RL model, is softened by the presence of a few long-range connections which decreases the path length and brings the average clustering coefficient closer to that of the data (see Figure \ref{fig:RL}B). The non-periodic boundary conditions lead to a small subpopulation of nodes with low degree. Because these nodes are neighbors in the adjacency matrix, they tend to be connected to one another, leading to an assortative topology. The same explanation underlies the existence of a hierarchical topology, because these low degree boundary nodes predominantly connect with one another.

\begin{figure}[h!]
\begin{center}
\includegraphics[width=0.66\textwidth]{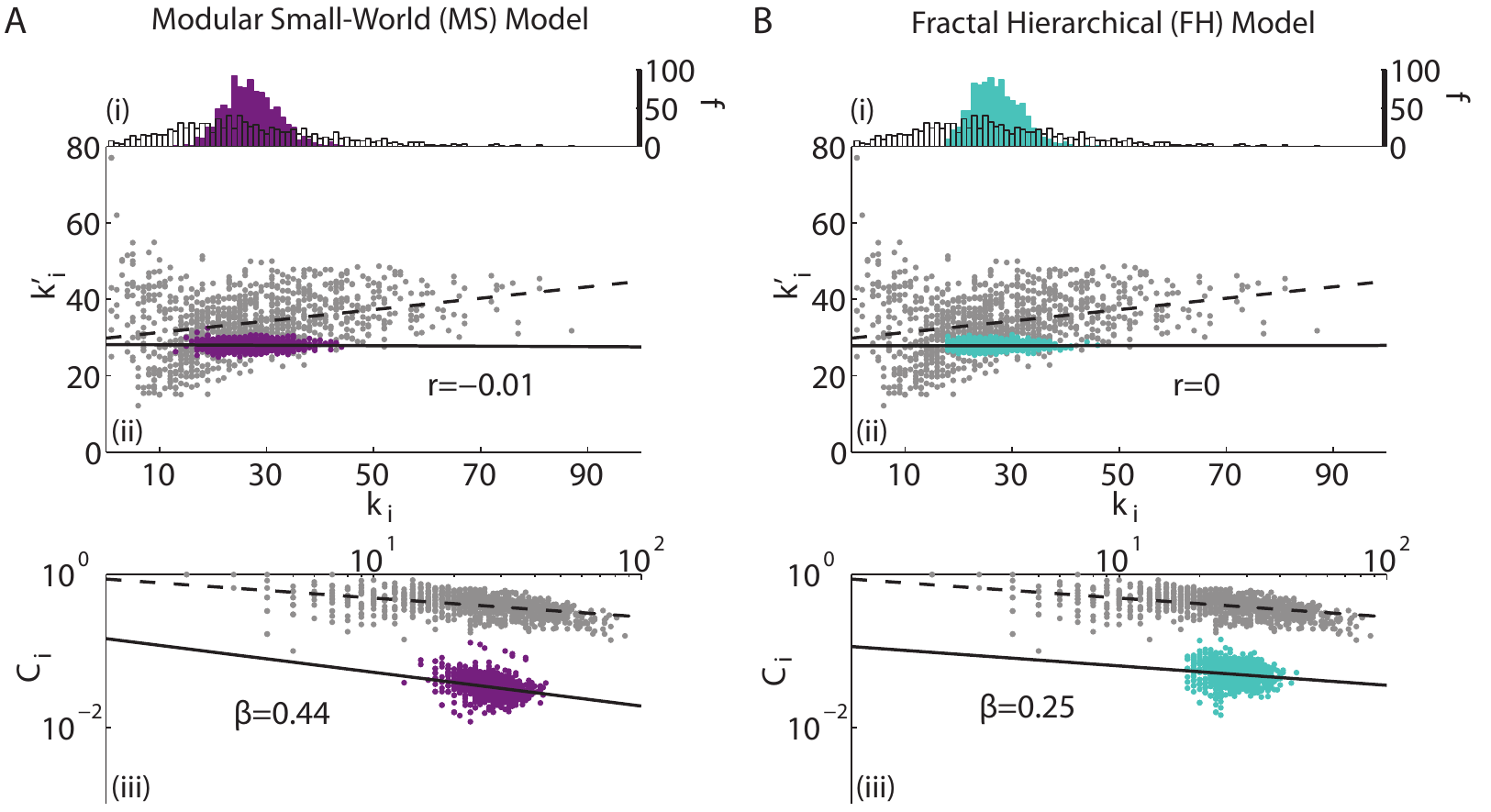}
\caption{Comparison between the \emph{(i)} degree distribution (number $f$ of nodes with a given degree $k_{i}$), \emph{(ii)} assortativity (correlation between a node's degree $k_{i}$ and the mean degree of that node's neighbors $k'_{i}$, summarized by parameter $r$), and \emph{(iii)} hierarchy (the relationship between the clustering coefficient $C_{i}$ and the degree $k_{i}$ over all nodes in the network, summarized by parameter $\beta$) of the \emph{(A)} modular small-world and the \emph{(B)} fractal hierarchical models and the same diagnostics in the brain anatomical data (grey). Black lines indicate best linear fit to the data (dashed) and model (solid) networks.} \label{fig:SW}
\end{center}
\end{figure}

\textbf{Modular Small-World (MS) Model:} Small world networks have received a great deal of attention \cite{Watts1998} as a conceptual characterization of structure that combines local order with long range connections. While the small world concept is sufficiently general that most networks that are not strictly regular or random fall into this category, small world organization represents more biologically relevant organization than the previous four cases \cite{Sporns2010,Bullmore2009,Bassett2010,Bassett2010c}. In addition to the small-world feature, biological networks including those extracted from human brain connectome data \cite{Meunier2009,Bassett2010c,Bassett2011b,Bassett2013} also often display community structure where set of nodes (modules) tend to be highly and mutually interconnected with one another combined with some long-distance connections.

For this study, we construct a synthetic small world network that consists of small, fully-connected modules composed of 4 nodes, randomly linked with one another with enough edges to match the density of the empirical network. This topology leads to high clustering, short path length, and small diameter \cite{Rubinov2009}. The randomly distributed inter-module links emanating from relatively high degree nodes decrease the clustering coefficient of these nodes because nodes in two different modules are unlikely to be otherwise linked. This structure therefore leads to a hierarchical topology (see Figure \ref{fig:SW}A(iii)). However, because the inter-module links are randomly distributed, nodes that contain such links are no more likely to share an edge with another such node than they are to share a link with any other node in the network. The model therefore does not display any observable assortativity (see Figure \ref{fig:SW}A(ii)).

\textbf{Fractal Hierarchical (FH) Model:} Like small world networks, fractal hierarchical topology has become a popular classification of networks and applies broadly, at least to some extent, to topologies that are neither regular nor random. Fractal hierarchical structure has been linked to some observed network structure in the brain \cite{Bassett2010c,Gallos2012,Meunier2009,Meunier2010,Zhou2006} and its use in neural network models produces several behaviors reminiscent of empirical neurobiological phenomena \cite{Wang2011,Kaiser2010,Rubinov2011b}.

To construct a fractal hierarchical model \cite{Ravasz2003}, we follow the approach outlined in \cite{Sporns2006}. We begin with a set of 4-node modules. We connect pairs of these 4-node modules with a probability $p_{1}$ to form 8-node modules. We connect pairs of 8-node modules with a probability $p_{2}$ to form 16-node module. Importantly, the probability $p$ of inter-module connections decreases at each level at a prescribed drop-off rate; that is, $p_{1}$ is larger than $p_{2}$, $p_{2}$ is larger than $p_{3}$, etc. The probabilities at each level are related to one another by a probability drop-off rate. This module-pairing process is repeated until we have formed a 1024-node fractal hierarchical network. To obtain a $N=998$ network comparable to the empirical brain data, we chose 26 nodes uniformly at random to delete from the network. If the network contained more (fewer) edges than the empirical network, we repeated the process with an increased (decreased) probability drop-off rate. The algorithm terminates when we obtain a network with the correct number of edges.

The fractal hierarchal network yields extremely similar results to the small world network in terms of the degree distribution, assortativity, and hierarchy (compare Figure \ref{fig:SW}A with Figure \ref{fig:SW}B). The striking similarities are surprising given the differences in how the two networks are constructed. While the networks share strong 4-node module building blocks, they differ in their coarser structure. The similarity in the results depicted in Figure \ref{fig:SW} suggest that the level-dependent structure in the fractal hierarchical model is not well-captured by these graph properties. Other types of network properties that specifically test for multiresolution phenomenon in brain structure might more readily distinguish between these two synthetic models \cite{Onnela2012,Lohse2013}.

\begin{figure}[h!]
\begin{center}
\includegraphics[width=0.66\textwidth]{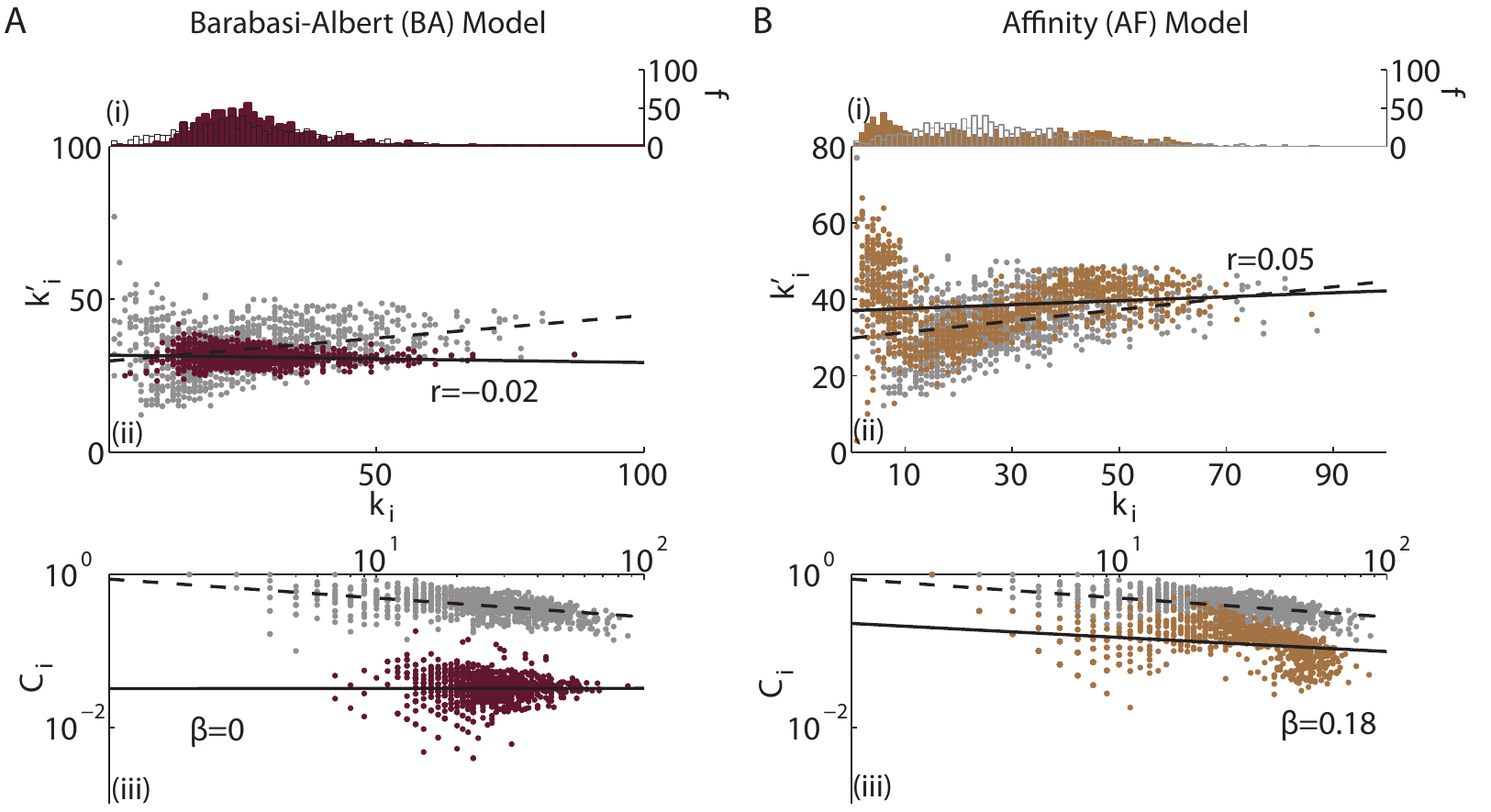}
\caption{Comparison between the \emph{(i)} degree distribution (number $f$ of nodes with a given degree $k_{i}$), \emph{(ii)} assortativity (correlation between a node's degree $k_{i}$ and the mean degree of that node's neighbors $k'_{i}$, summarized by parameter $r$), and \emph{(iii)} hierarchy (the relationship between the clustering coefficient $C_{i}$ and the degree $k_{i}$ over all nodes in the network, summarized by parameter $\beta$) of the \emph{(A)} Barab\'{a}si-Albert and \emph{(B)} affinity models and the same diagnostics in the brain anatomical data (grey). Black lines indicate best linear fit to the data (dashed) and model (solid) networks. In panel \emph{(B)}, the parameter values used for the affinity model are the following: $\gamma=1.94$, $\delta=3.48$, and $\epsilon=3.36$. } \label{fig:BA}
\end{center}
\end{figure}

\subsubsection{Growing Non-Embedded Models}

In this section we explore two non-embedded growth models (see Figure \ref{Fig1}). The first is the Barab\'{a}si-Albert preferential attachment model and the second is an affinity model which we design to capture assortative and hierarchical structure.

\textbf{Barab\'{a}si-Albert (BA) Model:} All models described thus far, with the exception of the configuration model, share a common and critical short-coming: the degree distribution is much narrower than that of the empirical networks. A model that produces a broader distribution of node degrees is the Barab\'{a}si-Albert model of preferential attachment \cite{BA2002}.

To construct a BA network, we begin with a single edge connecting two nodes. Then we iteratively add a single node to the network by linking the new node to $m$ existing nodes. The probability of linking the new node to an existing node is given by a preferential attachment function $\Pi(k)=k+k_0$ with dimensionless parameter $k_0$ tuning the rate of decrease in the degree distribution. Note that as $k_0 \rightarrow \infty$, the resultant graph becomes increasingly similar to an ER graph.

To identify a model network in this family that best fits the empirical data, we tune $k_0$ to minimize the difference between the model topology and the empirical topology as described in Section~\ref{sec:methods}. We find that networks constructed using $k_0=4$ provide the best available fit to the empirical data. The number of edges $m$ added with each new node is determined by the total number of edges $M$. This procedure produces networks with low clustering and broad degree distributions, although the number of low-degree nodes is underestimated in comparison to the empirical data (see Figure \ref{fig:BA}A(i)). Despite the broad degree distribution, the network does not display an assortative or hierarchical topology (see Figure \ref{fig:BA}A(ii)--(iii)).

\textbf{Affinity (AF) Model:} We introduce an extension of the BA model that includes constraints specifically designed to capture assortative and hierarchical structure. We define the affinity model by a two step preferential attachment function that does not depend on a node's current degree but instead depends on a dimensionless \emph{affinity} parameter $\alpha$. We begin with $N$ nodes, and to each node we assign a unique affinity $\alpha_{i}$ distributed uniformly at random in the interval [0,1]. The value of $\alpha_i$ remains unchanged throughout the growth process (see Algorithm \ref{alg:affinity}). We choose a node with probability $\propto \alpha_i^{\gamma}$ and link that node preferentially to another node $j$ with a similar affinity $\alpha_j$. This assortative mixing for affinity ensures degree assortativity. In addition, we choose a preferential attachment function (see Algorithm \ref{alg:affinity}, line 6) such that nodes with small values of affinity (e.g. small degree) are relatively more likely to gain edges with neighbors of similar affinity (and therefore degree) than nodes with large values of affinity. Small degree nodes therefore are more clustered than their high degree counterparts, leading to a hierarchical network structure.

\begin{algorithm}
 \SetAlgoLined
  \SetKwInOut{Input}{input}\SetKwInOut{Output}{output}
 \Input{number of nodes $N$\\ \ number of edges $M$\\ \ number of seed edges $M_0$\\ \ attachment regulators $\gamma$, $\delta$ and $\epsilon$}
 \Output{Adjacency matrix $A$}
  \BlankLine
initialize graph with $N$ nodes\;
connect $M_0$ pairs of nodes chosen uniformly at random\;
assign each node an affinity given by $\alpha_i=\frac{i-1}{N-1}$\;
 \While{$M'\mathrm{=current\ \# \ of\ edges}< M$}{
 out of the set of nodes with $k>0$ , choose a node $i$ with probability $\propto \alpha_i^{\gamma}$\\
 connect node $i$ to node $j$ (chosen at uniformly at random) with probability $\propto |\alpha_i -\alpha_j|^{\min \{0,-\delta + \epsilon \cdot \alpha_i\}}$    }
 \caption{Growth algorithm for the affinity model.} \label{alg:affinity}
\end{algorithm}

To compare this model to the empirical data, we use a derivative-free optimization method to identify the parameter values for $\gamma$, $\delta$, and $\epsilon$ that minimize the difference between the empirical and model networks; see Section~\ref{sec:methods}. The affinity model has a very broad degree distribution with a concentration of low degree nodes and an extremely heavy tail of high degree nodes (see Figure \ref{fig:BA}B(i)). The network is both assortative and hierarchical although the average clustering is lower than that found in the empirical data (see Figure \ref{fig:BA}B(ii)--(iii)). The randomly chosen edges connecting nodes of high degree induce a small diameter and short path length.

It is not surprising that the affinity model provides a better fit for the empirical data for these specific diagnostics than other synthetic networks we have considered so far, since it was specifically constructed to do so. This is, however, no guarantee that this algorithm will capture other network properties of the empirical data. Indeed, the fact that the affinity model also shows a similar topological dimension to the empirical brain network is surprising and interesting (see next section).

\subsubsection{Diagnostics Estimating the Topological Dimension:}

In this section, we compare topological measures of the empirical data with the set of 8 non-embedded synthetic networks: 6 static models and 2 growth models.

\begin{figure*}[b!]
\begin{center}
\includegraphics[width=1\textwidth]{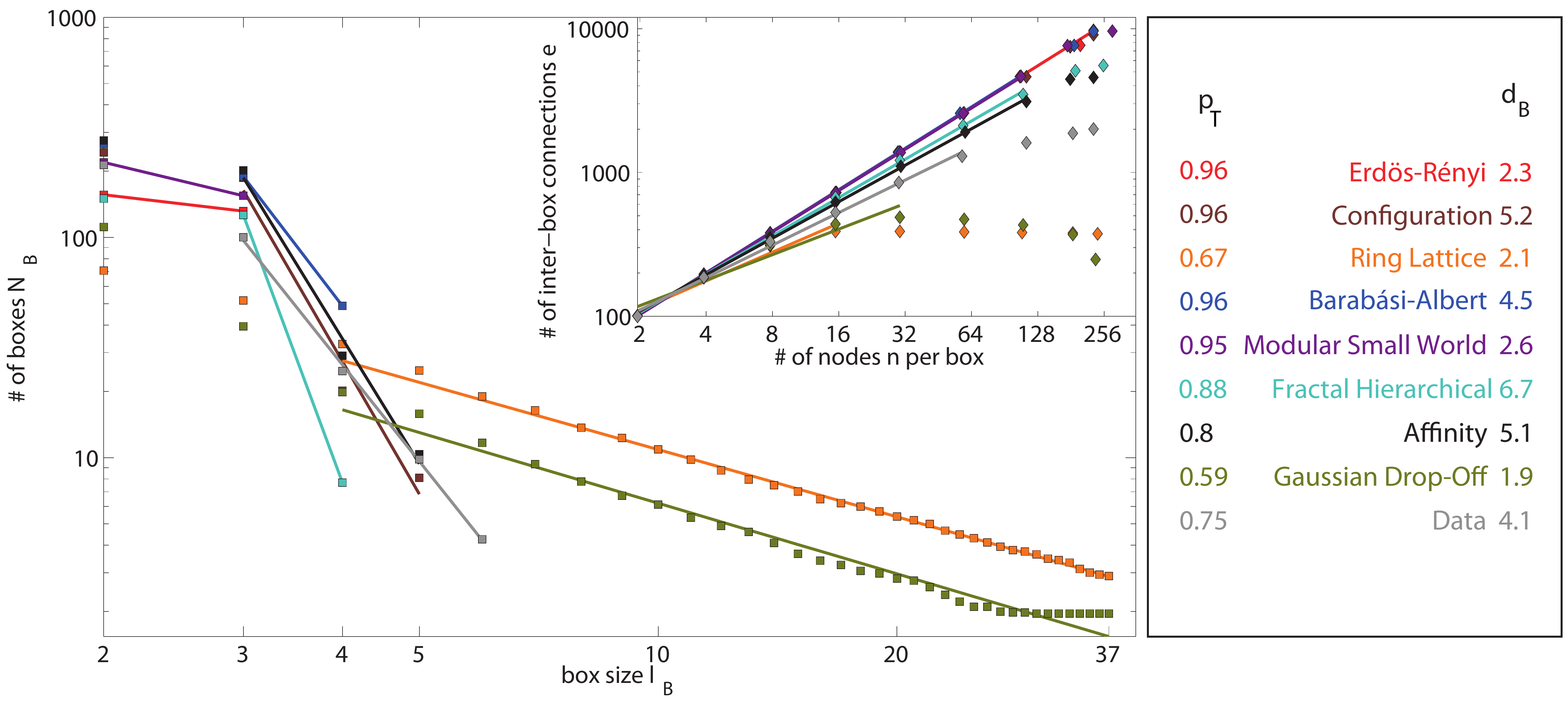}
\caption{\textbf{Diagnostics Estimating the Topological Dimension.} \emph{(Main Panel)} The number of boxes as a function of the topological size of the box, as estimated using the box-counting method \cite{Concas2006} (see Section~\ref{sec:methods}) for the real and synthetic networks. \emph{(Inset)} The topological Rentian scaling relationship between the number of edges crossing the boundary of a topological box and the number of nodes inside of the box (see Section~\ref{sec:methods}) for the real and synthetic networks. Lines indicate data points included in fits reported in Table \ref{tb:metrics}.} \label{fig:fracdim}
\end{center}
\end{figure*}

Using a box-counting method, we estimate the fractal dimension of the empirical and synthetic model networks (see Section~\ref{sec:methods}) and observe three distinct classes of graphs (see Figure \ref{fig:fracdim}, main panel). The first group, which includes the Erd\H{o}s-R\'{e}nyi and modular small-world models, has a diameter that is too small to allow an adequate estimation of the fractal dimension of the network using the box-counting method. The second group, which includes the Gaussian drop-off and ring lattice models, has a large diameter leading to a small fractal dimension. The third group, which includes the remainder of the models, has a similar diameter to the empirical network and therefore similar fractal dimension. By these comparisons, the affinity model is the best fit to the data and the configuration model is the second best fit.

The Gaussian drop-off and ring lattice models also show distinct topological Rentian scaling in comparison to the other models (see Figure \ref{fig:fracdim}, inset). Above a topological box size of 16 nodes, the number of inter-box connections does not increase because the edges are highly localized topologically. All other models display a swifter scaling of the number of edges with the number of nodes in a topological box in comparison to the empirical data. The affinity model displays the most similar scaling to that observed in the empirical data.

\clearpage

\subsection{Embedded Network Models}
The non-embedded models described in the previous section necessarily ignore a fundamental property of the brain: its embedding in physical space. Spatial constraints likely play an important role in determining the topological properties of brain graphs \cite{Bullmore2011,Bullmore2012,Bassett2010,Kaiser2006,Chen2006}. In this section, we explore the topological properties of spatially \textit{embedded graphs} in which the probability of connecting any two nodes in the network depends on the Euclidean distance between them \cite{Barthelemy2010}. We explore the same topological diagnostics as we did in the previous section: degree  distribution, assortativity, hierarchy, and diagnostics estimating the topological dimension of the network. As a whole, we find that spatially embedded models capture more topological features of the empirical networks than models that lack the physical embedding constraint.

\subsubsection{Static Embedded Models}

\textbf{Random Geometric (RG) Model:}
A random geometric model can be constructed by distributing nodes uniformly at random in a volume \cite{Barthelemy2010,Penrose2003,Dall2002}. We employ a classical neurophysiological embedding in which the x-axis represents the right-left dimension, the y-axis represents the anterior-posterior dimension, and the z-axis represents the superior-inferior dimension. We use a rectangular volume where the length of each side is equal to the maximal Euclidean distance between nodes as measured along that axis and we distribute $N$ nodes uniformly at random within this volume. The $M$ pairs of nodes with the shortest between-node distance are each connected by an edge.

\begin{figure}[h!]
\begin{center}
\includegraphics[width=1\textwidth]{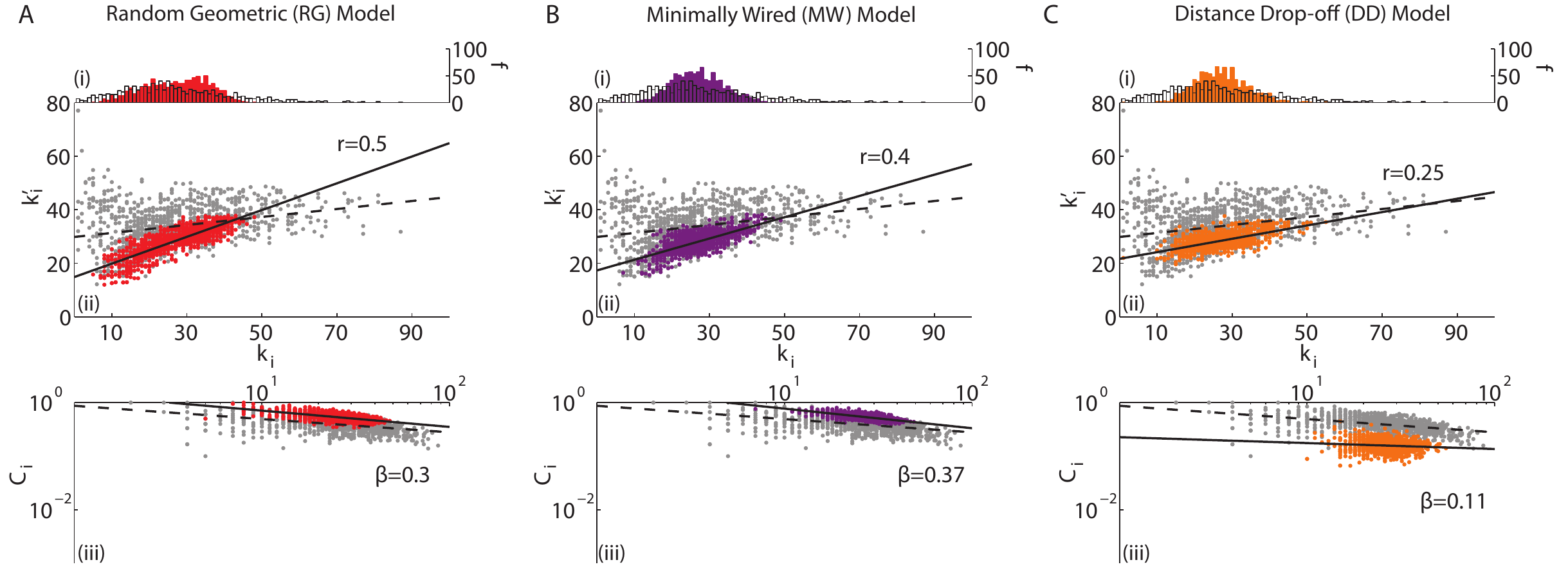}
\caption{Comparison between the \emph{(i)} degree distribution (number $f$ of nodes with a given degree $k_{i}$), \emph{(ii)} assortativity (correlation between a node's degree $k_{i}$ and the mean degree of that node's neighbors $k'_{i}$, summarized by parameter $r$), and \emph{(iii)} hierarchy (the relationship between the clustering coefficient $C_{i}$ and the degree $k_{i}$ over all nodes in the network, summarized by parameter $\beta$) of the \emph{(A)} random geometric (RG), \emph{(B)} minimally wired (MW), and \emph{(C)} distance drop-off (DD) models and the same diagnostics in the brain anatomical data (grey). Black lines indicate best linear fit to the data (dashed) and model (solid) networks.} \label{fig:RGG}
\end{center}
\end{figure}

The heterogeneity of node placement in the volume leads to a broad degree distribution and high clustering between spatially neighboring nodes, leading to a large network diameter and long path length (see Figure \ref{fig:RGG}A(i) and Table \ref{tb:metrics}). Because of the homogeneity of the connection rule, which is identical across all nodes, nodes with high degree (those in close proximity to other nodes) tend to connect to other nodes of high degree and nodes of low degree (those far from other nodes) tend to connect to other low degree nodes, leading to degree assortativity (see Figure \ref{fig:RGG}A(ii)). Nodes at the edges of spatial clusters will tend to have high degree but low clustering, leading to a hierarchical topology (see Figure \ref{fig:RGG}A(iii)).

\textbf{Minimally Wired (MW) Model:} As noted above, nodes in the RG model are placed uniformly at random in a volume. To add additional anatomical constraints to the model, we can construct a minimally wired model (MW) in which nodes are placed at the center of mass of anatomical brain regions. The $M$ pairs of nodes with the shortest between-node distance are then each connected by an edge.

The MW provides an interesting point of comparison to the RG because it allows us to assess what topological properties are driven by the precise spatial locations of brain regions alone. The degree distribution in the MW is narrower than it is in either the RG or the empirical brain network, likely because the brain parcellation used in this study is largely grid-like over the cortex (see Figure \ref{fig:RGG}B(i)). Like the RG, the MW displays degree assortativity and a hierarchical topology (see Figure \ref{fig:RGG}B(ii)--(iii)), and has high clustering and long path length. However, in general the diagnostic relationships extracted from the MW model do not match those of the empirical brain network as well as those extracted from the RG model.

\textbf{Distance Drop-Off (DD) Model:} Both the minimally wired and the random geometric models connect only the $M$ pairs of nodes with the shortest inter-node distance. These models therefore lack long distance connections which are known to be present in the brain, and have been argued to enable swift communication between distant brain areas \cite{Sporns2010}. To include this additional biological characteristic, we next study the distance drop-off model (DD) \cite{Avin2008}, in which we place nodes at empirical brain region locations and then connect pairs of nodes with a probability that depends on the distance $r$ between nodes: $P \propto g(r)$. Note that the minimally wired model is a special case of the DD model if we choose $P \propto g(r)$ to be a step function with threshold $r_0$. Here, however, we fit a function $g(r)$ to the connection probability of the empirical data as a function of distance (see Supplementary Material).

The results of the DD model are similar to those that we observed in the case of the minimally wired and random geometric models (see Figure \ref{fig:RGG}C). However, longer distance connections are present in this model which decrease the clustering, path length, diameter, and strength of the assortativity and hierarchy. In general, the diagnostic relationships extracted from the DD model match those of the empirical brain network significantly better than the same diagnostics extracted from the RG and MW models.

\subsubsection{Embedded Growth Models}

\textbf{Distance Drop-Off Growth (DDG) Model:} The random geometric, minimally wired, and distance drop-off models all have narrower degree distributions than the empirical data. To expand the degree distribution while still utilizing the empirical node placement and empirically derived probability function $P \propto g(r)$, we construct a distance drop-off growth model (DDG). We begin with $M_0$ seed edges which we distribute uniformly at random throughout the network. We then choose a node $i$ with $k_i>0$ uniformly at random and create an edge between node $i$ and node $j$ (with no constraint on $k_j$) according to the probability $P \propto g(r )$. We continue adding edges in this manner until the number of edges in the network is equal to $M$.

\begin{figure}[h!]
\begin{center}
\includegraphics[width=0.66\textwidth]{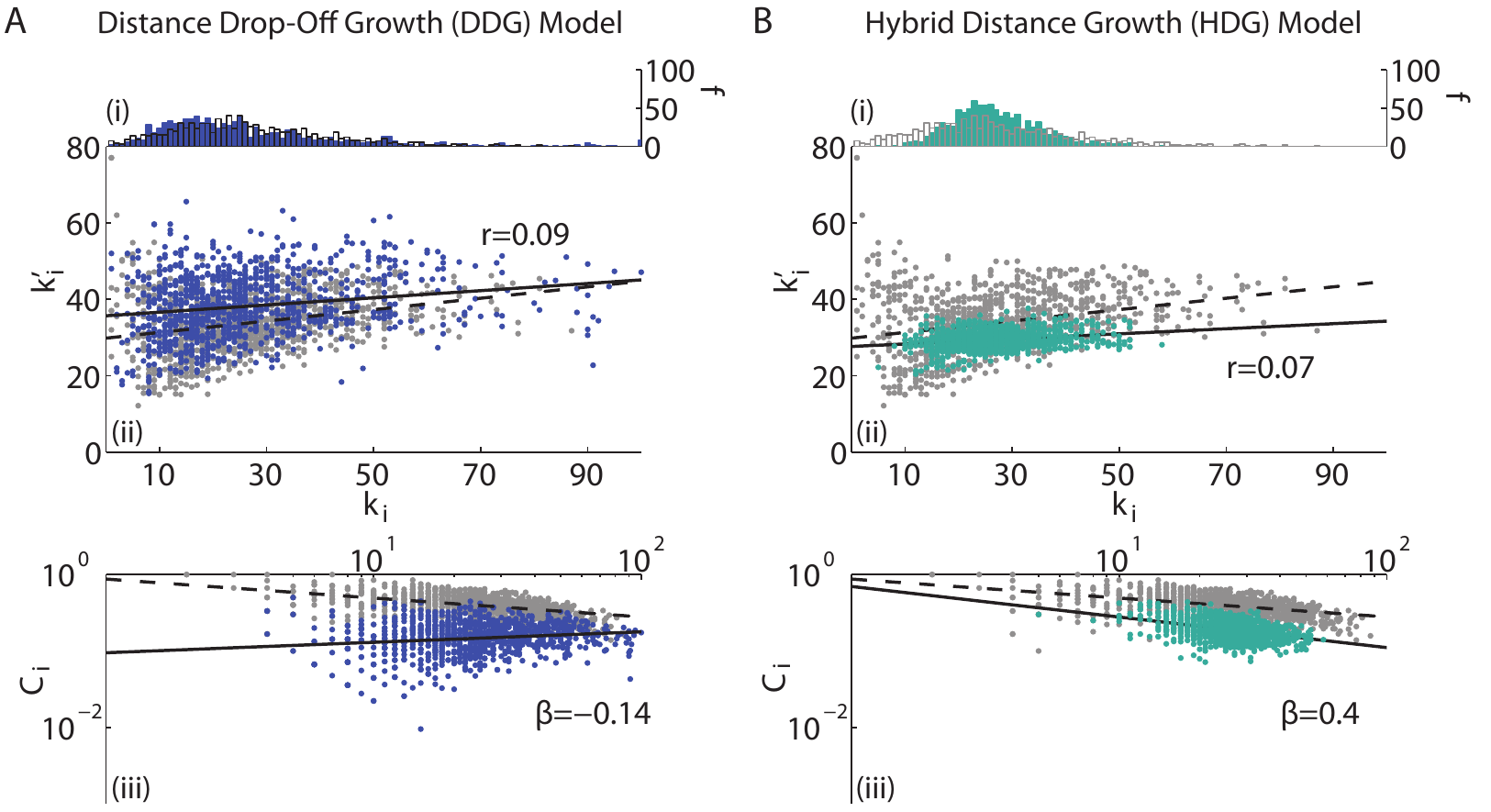}
\caption{Comparison between the \emph{(i)} degree distribution (number $f$ of nodes with a given degree $k_{i}$), \emph{(ii)} assortativity (correlation between a node's degree $k_{i}$ and the mean degree of that node's neighbors $k'_{i}$, summarized by parameter $r$), and \emph{(iii)} hierarchy (the relationship between the clustering coefficient $C_{i}$ and the degree $k_{i}$ over all nodes in the network, summarized by parameter $\beta$) of the \emph{(A)} distance drop-off growth (DDG) and the \emph{(B)} hybrid distance growth (HDG) models and the same diagnostics in the brain anatomical data (grey). Black lines indicate best linear fit to the data (dashed) and model (solid) networks. In panel \emph{(B)}, we use $4000$ minimized wired seed edges. } \label{fig:GRDG}
\end{center}
\end{figure}

The degree distribution and assortativity of the DDG are surprisingly similar to that observed in the empirical data (see Figure \ref{fig:GRDG}A(i)--(ii)). However, the stochasticity of the growth rule induces a decrease in clustering and we do not observe a hierarchical topology (see Figure \ref{fig:GRDG}A(iii)). Neither the network diameter nor the path length are significantly altered in comparison to the non-growing distance drop-off model.

\textbf{Hybrid Distance Growth (HDG) Model:} The values for all summary diagnostics reported in Table \ref{tb:metrics} for the above models that are most similar to the data are those calculated for the minimally wired and distance drop-off growth models. In a final model, we combine facets of both models in a hybrid distance growth model (HDG). We begin by creating a minimally wired model for the $M_0$ shortest connections. We then use the growing rule of the distance drop-off growth model to add the remaining $M-M_0$ edges to the network. This process can be interpreted as the creation of strongly connected functional modules that afterwards are cross-connected and embedded in the full network. Using a derivative-free optimization method, we estimate that the value of $M_0$ that produces a network most similar to the empirical network is $M_0 = 4000$; see Section~\ref{sec:methods}.

As expected, this hybrid model produces a degree distribution, assortativity, and hierarchy in between those produced by the minimally wired and distance drop-off growth models and therefore similar to those observed in the data (see Figure \ref{fig:GRDG}B(i)--(iii)). However, the clustering, diameter, and path length remain low in comparison to the empirical data (see Table \ref{tb:metrics}), suggesting that this model does not contain as much local order as the brain.

\subsubsection{Diagnostics Estimating the Topological Dimension}

In this section, we compare topological measures of the empirical data with the set of 5 embedded synthetic networks: 3 static models and 2 growth models.

\begin{figure*}[b!]
\begin{center}
\includegraphics[width=1\textwidth]{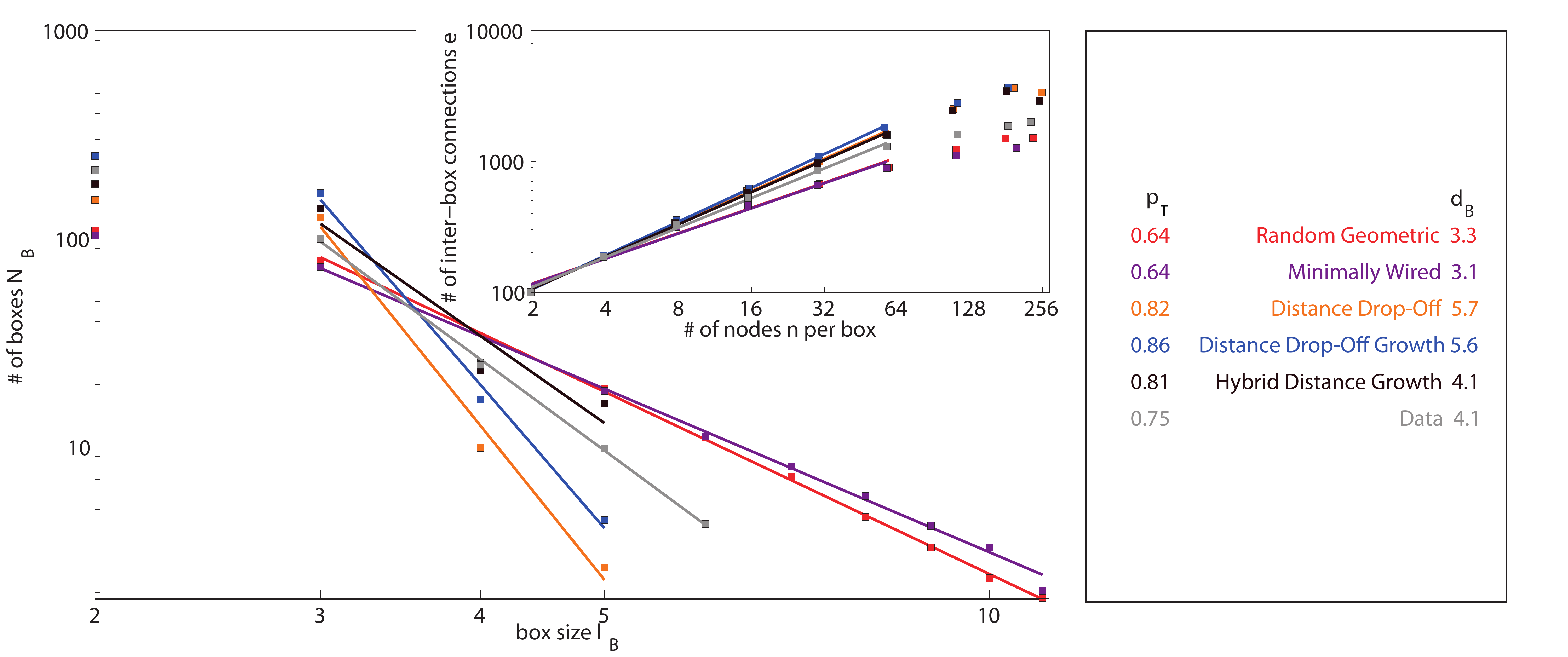}
\caption{\textbf{Diagnostics Estimating the Topological Dimension.} \emph{(Main Panel)} The number of boxes as a function of the topological size of the box, estimated using the box-counting method \cite{Concas2006} (see Section~\ref{sec:methods}) for the real and embedded model networks. \emph{(Inset)} The topological Rentian scaling relationship between the number of edges crossing the boundary of a topological box and the number of nodes inside of the box (see Section~\ref{sec:methods}) for the real and embedded model networks. Lines indicate data points included in fits reported in Table \ref{tb:metrics}.} \label{fig:fracdimphys}
\end{center}
\end{figure*}

We observe that the estimates of the topological dimension, using both box-counting and Rentian scaling methods, derived from the physical network models are more similar to the empirical data than those derived from the topological network models (see Figures~\ref{fig:fracdim} and~\ref{fig:fracdimphys}). The two highly locally clustered networks (the minimally wired and random geometric models) have larger diameters than the brain, decreasing their estimated fractal dimension in comparison. The distance drop-off and distance drop-off growth models are higher dimensional than the empirical data while the hybrid distance growth model displays the same dimension as the empirical data. The hybrid model also produces Rentian scaling with the most similar exponent to that obtained from the empirical data. The identified similarities between models and empirical data are somewhat surprising given that none of these models were explicitly constructed to attain a given topological dimension.

\newpage
\section{Discussion}
\addcontentsline{toc}{section}{Discussion}

We examined graph diagnostics of 13 synthetic network models and compared them to those extracted from empirically derived brain networks estimated from diffusion imaging data \cite{Hagmann2008}. We found that in general if a model was hard-coded to display one topological property of the brain (e.g., the degree distribution or the assortativity), it was unlikely to also display a second topological property, suggesting that a single mechanism is unlikely to account for the complexity of real brain network topology. We also observed that those models that employed information about node location and inter-node distances (e.g., embedded network models) were more likely to display similar topological properties to the empirical data than those that were constructed based on topological rules alone (e.g., non-embedded network models). In our examination, three models performed noticeably better than all others: the hybrid distance growing model, the affinity model, and the distance drop-off model. Together, these results provide us with important insights into the relationships between multiple topological network properties. Moreover, these model networks form a catalogue of null tests with a range of biological realism that can be used for statistical inference in static as opposed to dynamic network investigations \cite{Bassett2011b,Bassett2012c}.

\begin{figure*}[b!]
\begin{center}
\includegraphics[width=1\textwidth]{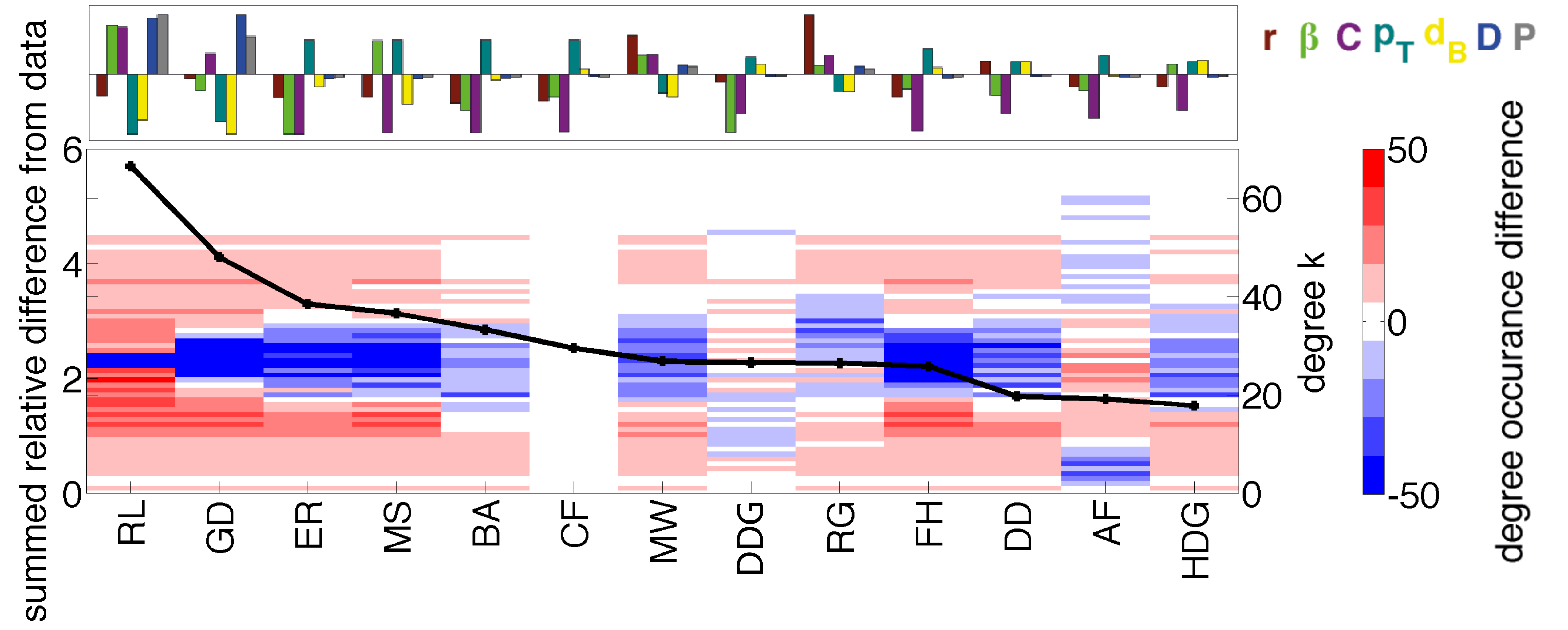}
\caption{\textbf{Comparison of the Network Models and Brain Data.} \emph{(Top Panel)} For each model, we illustrate how summary network statistics (Assortativity $r$, hierarchy $\beta$, clustering $C$, Rentian scaling $p_{T}$, fractal dimension $d_B$, diameter $D$, mean path length $P$) differ from the same statistics extracted from empirical data. \emph{(Main Panel)} The black line indicates the sum of the absolute values of the relative difference between each model and the data. The color image in the background indicates the difference between the degree distribution of the model and that of the data: red colors indicate that the model has too many nodes of a given degree, while blue colors indicate that the model has too few nodes of a given degree. Less saturated colors indicate more similarity between the degree distributions of the model and the data. } \label{fig:compare}
\end{center}
\end{figure*}

Figure~\ref{fig:compare} provides a summary of graph diagnostics extracted from real and synthetic model data. We measure the relative difference between model and data, normalized by the value obtained from the model that fits the data the least for each diagnostic: $(r_{model}-r_{data})/\max \{r_{all\ models}\}$. Models are placed in descending order, from those with the largest relative difference to the data (left-most side of the graph) to those with the smallest relative difference to the data (right-most side of the graph). We observe that embedded models generally have a smaller relative distance to the empirical data than non-embedded models. This result supports the view that physical constraints likely play an important role in large-scale properties of neurodevelopment.

\subsection{Non-embedded Models}
We probe non-embedded models with differing amounts and types of structure. While the Erd\H{o}s-R\'{e}nyi model provides an important benchmark with a random topology, it bears little resemblance to the brain network. Although a homogeneous random distribution of links has been suggested to characterize the small-scale structure of neuron-to-neuron connections \cite{Braitenberg1998,Henderson2011}, the large-scale structure of human and animal brains instead displays heterogeneous connectivity \cite{Sporns2010}. Perhaps one of the simplest measures of this heterogeneity is found in the degree distribution, which displays a predominance of low degree nodes and a long tail of high degree nodes. In comparing the degree distribution of the brain to that obtained from a BA model, it is clear that this tail, however, is not well-fit by a power-law, a finding consistent with previous reports in brain anatomy \cite{Humphries2006,Bassett2008} and function \cite{vandenHeuvel2008,Achard2006}.  However, by matching the empirical data, for example using a configuration model with the same degree distribution, we note that we do not automatically uncover higher order structures like assortativity, suggesting that the degree distribution provides only limited insight into the forces constraining brain network development.

Several decades ago, neuroanatomists observed that the \emph{pattern} of connections in several animal brains displayed a combination of both densely clustered areas and long range projects between distant areas \cite{Scannell1995,Scannell1999,Young1995,Felleman1991}. The regular lattice and Gaussian drop-off models are able to capture these densely connected structures but fail to capture the extent of long-range connectivity observed in the brain. The small-world modular and fractal hierarchical models contain both properties: dense local connectivity and long-range interactions. The fractal hierarchical model has the added benefit of containing nested structures, which have been implicated in the heterogeneity of neuronal ensemble activity \cite{Rubinov2011b} and in the separation and integration of information processing across multiple frequency bands \cite{He2010}. Moreover, hierarchical modular structure has been identified in organization of white matter streamlines in human diffusion weighted imaging data \cite{Bassett2010c,Gallos2012,Zhou2006} and implicated in neurobiological phenomena \cite{Wang2011,Kaiser2010,Rubinov2011b}.

None of the non-embedded models discussed earlier in this section simultaneously provide a heterogeneous degree distribution, degree assortativity, hierarchical topology, and realistic topological dimensions. Such a ``No Free Lunch'' rule is perhaps unsurprising, in that a network that is developed to directly obtain one property typically fails to also display a second property. However, this result suggests that the topological properties that we explore here are in some sense independent from one another. In light of the previously reported correlations between network diagnostics in human brain networks \cite{Lynall2010} suggesting the need for methods to identify distinguishing properties \cite{Bounova2012,Onnela2012}, our finding interestingly suggests that such correlations are potentially specific to the brain system and not expected theoretically.

Finally, in our affinity model, we hard-code both degree assortativity and a continuous hierarchical topology, rather than the discrete hierarchy employed in nested models like the fractal hierarchical model examined here. Interestingly, however, and in contrast to the other non-embedded models, we simultaneously obtain a heterogeneous degree distribution, and similar estimates of the topological dimension. This model fits multiple properties of brain networks that were not explicitly included in the construction of the network model, but are nevertheless a consequence of a three-parameter fit in the specific affinity model selected. The affinity model therefore serves as a promising candidate as both a generative model and statistical null model of brain organization.

\subsection{Embedded Models}

In an effort to include additional biological constraints, we also explore several models that employ information regarding either the physical placement of network nodes or that place constraints on the Euclidean lengths of network edges. In general, this set of networks outperforms most of the non-embedded network models that we studied, supporting the notion that physical constraints might play important roles in brain network development and structure \cite{Kaiser2006,Chen2006,Bullmore2012,Vertes2012,Bassett2010,Bassett2010c,Raj2011,Henderson2011}.

It is important to preface the discussion of our results by mentioning the fact that the properties of empirically derived brain networks display a heterogeneity that could at least in part stem from the peculiar physical properties of the organ. Brains are symmetric objects, with the two hemispheres being connected with one another via tracts in the corpus callosum and via subcortical structures. This separation allows for a very different topology \emph{within} a hemisphere than \emph{between} hemispheres. Moreover, cortical areas (gray matter) form a shell around the outer edges of the brain while their connections (white matter) compose the inner volume. Finally, brain areas are inherently heterogeneous in physical volume, making their distances from one another far from homogeneous. While the morphology of the brain constrains its potential topological properties, evidence also suggests that the lengths of tracts connecting brain areas follow a heavy tailed distribution, with short tracts being relatively common and long tracts being relatively rare \cite{Kaiser2006,Chen2006}. These findings are in concert with the idea that energy efficiency---to develop, maintain, and use neuronal wiring---remains a critical factor in brain evolution and development \cite{Attwell2001,Bullmore2012}.

In this study, we begin with a random geometric model, whose nodes are placed uniformly at random in a volume but whose edges selectively link nodes that are nearby in physical space. In light of the simplicity of this model, it is somewhat surprising that we obtain such good agreement with the empirical degree distribution, the presence of assortativity, and the presence of a hierarchical topology. In the minimally wired graph we employ a similar connection rule but also fix node placement to be identical to that in the empirical brain network, following previous studies \cite{Bassett2010}. However, neither of these two models are able to capture the extent of long-distance connections in the empirical data. By employing the distance drop-off model, we can fix a connection \emph{probability} that varies with distance, rather than simply a connection \emph{threshold}. This connection probability, however, is not enough to provide a realistically broad degree distribution. Our distance drop-off growth model combines the strengths of each of these models by laying down a set of seed edges uniformly at random in a volume and then iteratively adding edges between pairs of nodes according to a probability that falls off with inter-node distance. The resulting degree distribution and assortativity properties are the best match to the empirical data of the models that we studied. A hybrid between the minimally wired model and the distance drop-off growth model does not perform significantly better in matching these properties and shows a hierarchical structure that is more pronounced than the data.

Importantly, the embedded network models examined here are purposely simplistic. While arbitrarily more complex models could be constructed, our goal was to isolate individual drivers of topology and probe their relationship to observed network diagnostics. Other studies of interest in relation to these findings include those that explore the effects of geometric folding \cite{Henderson2011}, radial surface architectures \cite{Raj2011}, and the effects of wiring minimization on functional networks \cite{Vertes2012}.

\subsection{Biological Interpretations}

While the construction of network models is genuinely critical in providing null tests for statistical inference of brain structure from data, this avenue of research also has the potential to provide key insights into the neurobiological mechanisms of brain development and function if performed with appropriate caution. In light of this second use, we note that several of the network models discussed in this paper employ rules that are reminiscent of---or even directly inspired by---known biological phenomena. For example, physical models that place constraints on the length of connections in Euclidean space are consistent with the known distribution of connection lengths in the brain and the modern understanding of metabolic constraints on the development, maintenance, and use of long wires \cite{Attwell2001,Bullmore2012,Kaiser2006,Chen2006,Bassett2010}.

However, even topological constraints that link nodes that have similar sets of neighbors can be interpreted as favoring links between neurons or regions that share similar excitatory input \cite{Vertes2012}. As an example, our affinity model hard-codes two inter-node relationships. First, nodes with a similar degree are more likely to be connected to one another by an edge, leading to degree assortativity throughout the network. This behavior can be thought of as a mathematical representation of the intuitive principle of spatial homophily: large neurons with expansive projections (e.g., pyramidal or basket cells) are more likely to connect to one another because they densely innervate tissue over large distances. Network assortativity can also stem from the temporal homophily that occurs during development: neurons that migrate over longer distances during development are more likely to come into contact with---and therefore generate a synapse with---one another than neurons that migrate over shorter distances. The second topological relationship hard-coded into the affinity model is the prevalence of clustering in local neighborhoods, a property consistent with physical constraints on network development. As neurons develop, it is intuitively more likely for them to create synapses with neighboring neurons than non-neighboring neurons, thereby closing topological loops in close geographic proximity. While we have only provided a few examples here, links between topological rules and biological phenomena provide potentially critical neurophysiological context for the development and assessment of synthetic network models.

\subsection{Future Directions}
The perspective that we have taken in choosing synthetic network models is one of parsimonious pragmatism. We seek to identify models with simplistic construction rules or growth mechanisms to isolate topological (non-embedded) and physical (embedded) drivers of network topology. One alternative perspective would be to begin with a certain graph topology (for example, an Erd\H{o}s-R\'{e}nyi graph), and iteratively rewire edges to maximize or minimize a network diagnostic or set of network diagnostics \cite{Vertes2012}. However, this approach requires prior hypotheses about which network diagnostics are most relevant for brain network development, a choice that is complicated by the observed correlations between such diagnostics \cite{Lynall2010}. Another approach is to employ exponential random graph models \cite{vanWijk2010,Simpson2011,Simpson2012}, which provide a means to generate ensembles of networks with a given set of network properties but do not provide a means to isolate mechanistic drivers of those network properties. A third approach is to construct a mechanistic model based on particle-particle collisions, which might serve as a physical analogy to the biological phenomena of neuronal migration through chemical gradients \cite{Gonzalez2006,Lind2007}. In each of these cases, a perennial question remains: at what spatial scale should we construct these models to gain the most insight into the relevant biology? Important future directions could include the development of multiscale growth models, enabling us to bridge the scales between neuronal mechanisms and large-scale structure.

\subsection{Methodological Limitations}
\addcontentsline{toc}{subsection}{Methodological Limitations}
There remain important limitations to our work. First, the development of high resolution imaging methods and robust tractography algorithms to resolve crossing fibers are fast-evolving areas of research. Novel imaging techniques have for example recently identified the existence of 90 degree turns in white matter tracts \cite{Wedeen2012}, a biological marker that we are not sensitive to in our data. Secondly, we have focused on understanding the (binary) topology of brain network architecture rather than its weighted connection strengths in this first study. Our choice was informed by three factors: 1) An understanding of the relationship between synthetic network models and brain network topology could be useful for informing a similar investigation into network geometry, 2) In these particular networks, node degree (binary) and node strength (weighted by the number of streamlines) are strongly correlated (Pearson's correlation coefficient $r=0.41$, $p=1 \times 10^{-41}$) and therefore topology serves as a proxy for weighted connectivity, and 3) The choice of how to weight the edges in an anatomical network derived from diffusion imaging is an open one \cite{Rubinov2011}, and therefore investigations independent of these choices are particularly useful. Finally, our models could be extended to include additional physical features of the human brain such as bilateral symmetry, the topological ramifications of which are not well understood.

\subsection{Conclusion}
\addcontentsline{toc}{subsection}{Conclusion}

In this paper, we have examined the mechanistic drivers of network topologies by employing and developing a range of synthetic network models governed by both topological (non-embedded) and physical (embedded) rules and comparing them to empirically derived brain networks. We hope that these tools will be useful in the statistical inference of anatomical brain network structure from neuroimaging data and that future work can build on these findings to identify neurobiologically relevant mechanisms for healthy brain architecture and its alteration in disease states.

\begin{sidewaystable}
{\tiny
\hfill{}
\hspace{-2cm}
\begin{tabular}{|c||c|c|c|c|c|c|c|c|c|}
\hline
\textbf{Network}&\begin{sideways}\textbf{mode degree (\#)}\end{sideways}& \begin{sideways}\textbf{width of degree distribution $\min{k_i}$--$\max{k_i}$}\end{sideways}& \begin{sideways}\textbf{assortativity $r$ }\end{sideways}&\begin{sideways}\textbf{hierarchy $\beta$}\end{sideways}&\begin{sideways}\textbf{global clustering $C$ $[\%]$}\end{sideways}&\begin{sideways}\textbf{Rentian scaling $p_{T}$}\end{sideways}&\begin{sideways}\textbf{topological fractal dimension $d_B$}\end{sideways} &\begin{sideways}\textbf{diameter $D$}\end{sideways} &\begin{sideways}\textbf{average shortest path length $P$}\end{sideways}\\ \hline
\hline
\emph{\textbf{Empirical}} & ~ & ~ & ~ & ~ & ~ & ~ & ~ & ~ & ~ \\ \hline
Brain& \textbf{23 \& 25 (40)}&\textbf{0--87}&\textbf{ 0.149}&\textbf{ 0.247} &\textbf{ 41.5}& $\mathbf{0.745\pm0.003}$ & $\mathbf{3.7\pm0.1}$ &\textbf{6}&$\mathbf{3.1}$\\ \hline
\emph{\textbf{Non-embedded}} & ~ & ~ & ~ & ~ & ~ & ~ & ~ & ~ & ~ \\ \hline
\emph{Static} & ~ & ~ & ~ & ~ & ~ & ~ & ~ & ~ & ~ \\ \hline
ER&$27\pm 1$ ($86 \pm 6$)&($12\pm2$)--($45\pm2$)&$-0.001 \pm 0.009$&$ -0.11 \pm 0.07$&$ 2.72 \pm 0.05$&$0.953 \pm 0.003$&$3 \pm 1$&$ 3.8 \pm 0.4$&$2.44 \pm 0.07$\\
CF&\textbf{23 \& 25 (40)}&($\mathbf{0}$)\textbf{--}($\mathbf{87}$)&$-0.02 \pm 0.01$&$0.11 \pm 0.03$&$4.32 \pm 0.08$&$ 0.952 \pm 0.005$&$4.0 \pm 0.2$&$ 5.0 \pm 0.1$&$2.4 \pm 0.2$\\
RL&$27$ ($470 \pm 20$)&($24.5 \pm 0.6$)--($28$)&$0.01 \pm 0.01$&$0.54 \pm 0.02$&$71.98 \pm 0.04$&$0.386 \pm 0.001$&$1.09 \pm 0.03$&$37.2\pm 0.4$&$18.82 \pm 0.09$\\
GD&$27.3\pm 0.8$ ($136\pm 8$)&($12\pm 1$)--($37\pm 1$)&$0.12 \pm 0.01$&$0.15 \pm 0.02$&$55.0\pm 0.2$& $0.463 \pm 0.003$&$0.3 \pm 0.1$&$38.7 \pm 0.6$&$13 \pm 3$\\
MS&$26 \pm 1$ ($91 \pm 5$)&($14 \pm 1$)--($44 \pm 2$)&$0.002 \pm 0.009$&$0.45 \pm 0.05$&$3.60 \pm 0.05$&$0.952 \pm 0.003$&$2 \pm 2$&$3.5 \pm 0.5$&$2.44 \pm 0.07$\\
FH&$26 \pm 1$ ($93 \pm 5$)&($17.8 \pm 0.5$)--($44\pm 2 $)&$0.002 \pm 0.008$&$0.16 \pm 0.07$&$5.1 \pm 0.1$&$0.898 \pm 0.002$&$4.1 \pm 0.1$&$4 $&$2.47 \pm 0.07$\\ \hline
\emph{Growth} & ~ & ~ & ~ & ~ & ~ & ~ & ~ & ~ & ~ \\ \hline
BA&$23\pm 2$ ($55\pm 4$)&($4\pm 2$)--($80\pm 9$)&$-0.035\pm 0.010$&$0.03\pm0.04$&$3.46\pm 0.09$ & $0.954\pm 0.004$ & $3.4\pm 0.7$ & $4.0 \pm 0.1$ & $2.4 \pm 0.1$\\
AF&$7\pm 6$ ($46 \pm 5$)&($1.0\pm 0.1$)--($73\pm 5$)&$0.07\pm 0.03$&$0.15\pm 0.08$&$13\pm 1$&$0.86 \pm 0.01$& $3.6 \pm 0.3$&$4.8\pm 0.4$&$2.5 \pm 0.3$   \\
\hline
\emph{\textbf{Embedded}} & ~ & ~ & ~ & ~ & ~ & ~ & ~ & ~ & ~ \\ \hline
\emph{Static} & ~ & ~ & ~ & ~ & ~ & ~ & ~ & ~ & ~ \\ \hline
RG&$29\pm 3$ ($51\pm 4$)&($4 \pm 1$)--($ 50\pm 3$)&$0.53\pm 0.03$&$0.30 \pm 0.03$&$53.9 \pm 0.4$&$0.644 \pm 0.008$ &$2.7 \pm 0.2$&$10.4 \pm 0.5$&$4.5 \pm 0.6$\\
MW&28 (65)&7--49&0.3971&0.3657&54.66&$0.635\pm 0.0026$&$2.4\pm 0.1$&11&$5.2 \pm 0.5$\\
DD&$26\pm 2$ ($68\pm 4$)&($1.3 \pm 0.6$)--($53\pm 3$)&$0.23 \pm 0.01$&$0.12 \pm 0.04$&$16.4\pm 0.2$&$0.819\pm 0.004$&$4.4 \pm 0.3$&$4.9 \pm 0.3$&$2.7 \pm 0.1$\\ \hline
\emph{Growth} & ~ & ~ & ~ & ~ & ~ & ~ & ~ & ~ & ~ \\ \hline
DDG&$\mathbf{16\pm 6}$ ($\mathbf{40\pm 5}$)&($2\pm 1$)--($231\pm 65$)&$0.1\pm 0.1$&$-0.10\pm 0.09$&$16.4 \pm 0.6$&$0.849\pm 0.006$ &$4.3 \pm 0.5$&$5.0\pm 0.5$&$2.6 \pm 0.2$\\
HDG & $24\pm 2$ ($55\pm 4$)& ($7\pm 1$)--($66\pm 6$) & $0.07 \pm 0.02$ & $0.31\pm 0.04$ & $17.7\pm 0.3$ & $0.821\pm 0.004$ & $4.5\pm 0.6$ & $4.6\pm 0.5$ & $2.6\pm 0.1$\\
\hline
\end{tabular}}
\hfill{}
\caption{Diagnostics of network topology in real and synthetic network models. The mean and standard deviation is reported for 100 realizations of each synthetic network model and the empirical value is reported for the brain data (see Supplementary Material for additional results for other participants). Note that the algorithms for determining fractal dimension and Rentian scaling are not deterministic and therefore multiple estimates can be obtained by applying the algorithms to the same network multiple times.}
\label{tb:metrics}
\end{sidewaystable}


\section*{Acknowledgments}
This work was supported by the Errett Fisher Foundation, Sage Center for the Study of the Mind, Fulbright Foundation, Templeton Foundation, David and Lucile Packard Foundation, PHS Grant NS44393, German Academic Exchange Service and the Institute for Collaborative Biotechnologies through contract no. W911NF-09-D-0001 from the U.S. Army Research Office, and the National Science Foundation (DMS-0645369). The funders had no role in study design, data collection and analysis, decision to publish, or preparation of the manuscript. We thank Olaf Sporns for giving us access to the human diffusion spectrum imaging data used in this work.

\newpage

\clearpage
\newpage

\section*{Supplementary Materials}

\subsection*{Extracting Empirical Connection Probability Drop-off}

In the distance drop-off (DD), distance drop-off growth (DDG), and hybrid distance growth (HDG) models we used a connection density $g$ that depends on the Euclidean distance $r$ between a pair of nodes. To tune our models, we estimated the empirical relationship between connection density and distance, hereafter referred to as \emph{connection probability drop-off}, from the brain data. We observed that the probability drop-off of connections that lay within a single hemisphere displayed significantly different behavior than the probability drop-off of connections that lay between hemispheres. We did not observe systematic differences between the connection probability drop-off in the two hemispheres, and therefore treat all within-hemispheric connections identically without distinguishing connections that lay in the left hemisphere from connections that lay in the right hemisphere.

To estimate the connection probability drop-off function $g(r)$ from the empirical data, we used an adaptive binning algorithm that determines $\frac{\mathrm{\#\  connections}}{\mathrm{\#\ possible\ connections}}$ for each bin width $\Delta _r$, where $\Delta_{r}$ is chosen to ensure that each bin contains on average 50 connections. To fit the data with sufficient precision, we first define a truncated power-law function using the form
\begin{equation}
\label{eq:connprob1}
f(x) = cx^{\alpha}\exp{(-\lambda x)}
\end{equation}
where $c$ is a constant, $x$ is the minimum physical distance of connections in the bin, $\alpha$ is the power-law exponent, and $\lambda$ is the natural exponent. However, we observed that this single truncated power-law fit was unable to adequately estimate the preponderance of long distance connections in the data. We therefore used a piecewise function with two truncated power-law functions of the form
\begin{equation}
\label{eq:connprob}
	g(x) =
	\begin{cases} f(x) &\mbox{if } x>x_0\\
	\frac{1}{2} f(x)+\frac{1}{2}f(x_0)\frac{x}{x_0}^{-\gamma}&\mbox{if } x<x_0
	\end{cases}
\end{equation}
where $f(x)$ is as defined in Equation 6, $\lambda$ is the power-law exponent for all bins in which $x<x_0$, $x_{0}$ is the minimum physical distance at which the truncated power-law function begins to fit the data, and $\gamma$ is the power-law exponent for all bins in which $x>x_0$.  To minimize boundary and resolution effects, we excluded the bin with the smallest minimum physical distance and the bin with the largest minimum physical distance.

\begin{figure*}[b!]
\begin{center}
\includegraphics[width=0.7\textwidth]{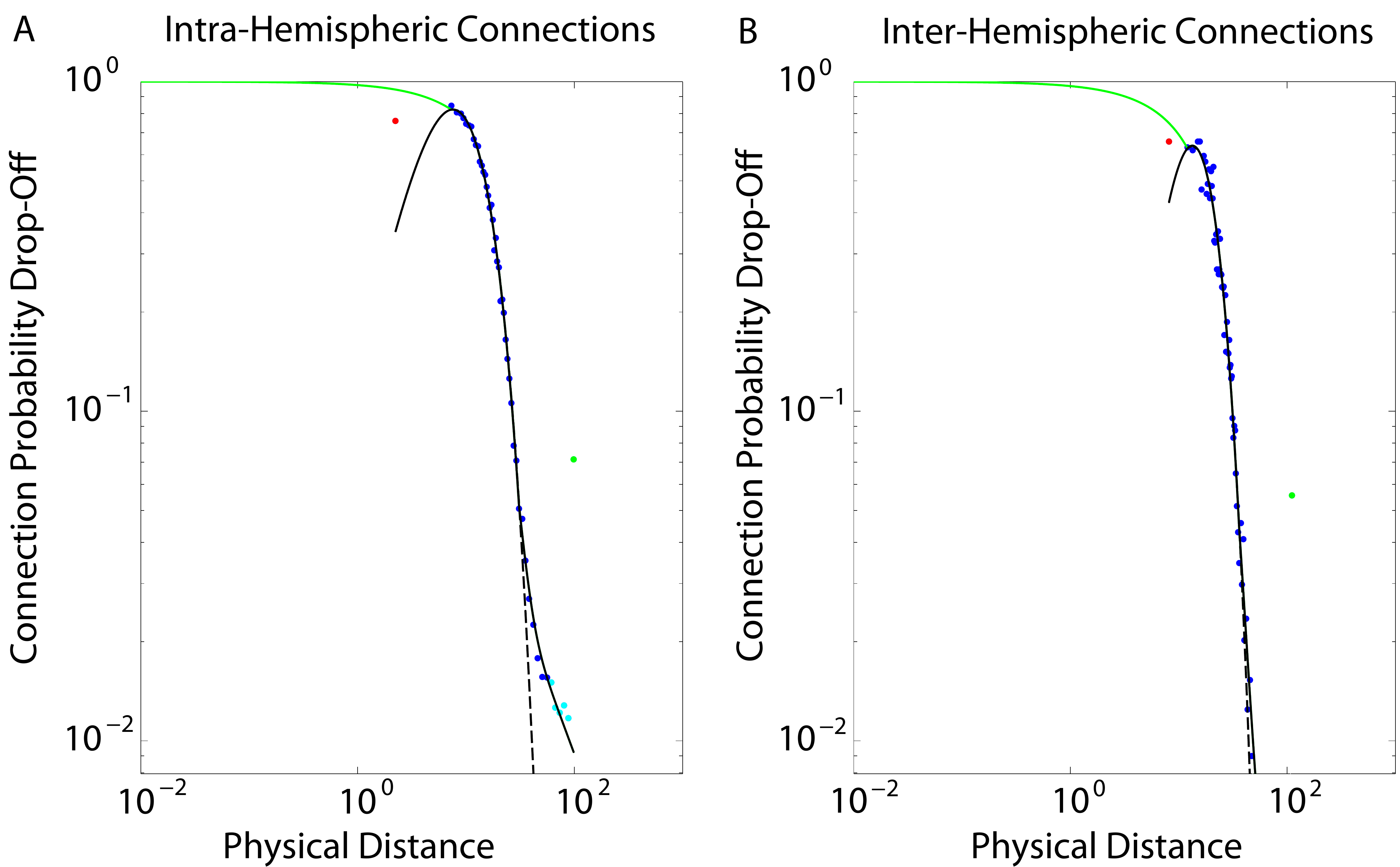}
\caption{\textbf{Empirical Connection Probability Drop-Off with Physical Distance} The connection probability drop-off $g(r)$ for \emph{(A)} intra- and \emph{(B)} inter-hemispheric connections. Empirical brain data is given by the data points: red indicates bins that were not utilized in the fits, blue indicates bins in which $x<x_0$, cyan indicates bins in which $x>x_0$, green indicates outlier bins excluded from fit. Fits are given by the lines: dotted line indicates the initial single truncated power-law fit, solid black line indicates the piecewise truncated power-law fit, and solid green indicates the piecewise truncated power-law fit with the interpolation to $g(0)=1$.}
\label{fig:connprob}
\end{center}
\end{figure*}

Algorithmically, we first fit a single truncated power-law function to the data to obtain estimates for the parameters $c$, $\alpha$, and $\lambda$ (see dashed line in Figure~\ref{fig:connprob}). We then used these estimates as initial parameters in the fit of Equation 7 to the data (see solid line in Figure~\ref{fig:connprob}). To obtain boundary values for the function, we performed a linear interpolation from the bin with the smallest minimum physical distance to the boundary value $g(0) = 1$ (see green line in Figure~\ref{fig:connprob}).

The results of this model, which contains 5 tunable parameters ($c$,$\alpha$,$\lambda$,$x_0$,$\gamma$), are shown in Figure \ref{fig:connprob}, where we observe a good agreement between the fit and the data. The estimated parameter values are provided in Table~\ref{Tab1}.

\begin{table}[h]
\begin{center}
\begin{tabular}{| l | l | l | l | l | l |}
\hline
Type             & $c$     & $\alpha$ & $\lambda$ & $x_0$      & $\gamma$ \\
\hline
Intra-Hemisphere & $0.1490$ & $1.6608$ & $0.2188$ & $30.8774$ & $0.8890$ \\
Inter-Hemisphere & $0.0033$ & $3.4492$ & $0.2695$ & $36.1939$ & $4.0198$ \\
\hline
\end{tabular}
\end{center}
\vspace{-5mm} \caption{\textbf{Parameter Estimates for Empirical Connection Density Drop-Off} for the fits of Equation 7 to intra- and inter-hemispheric data. \label{Tab1}}
\end{table}

\clearpage
\newpage

\begin{figure*}[b!]
\centering
\includegraphics[width=1\textwidth]{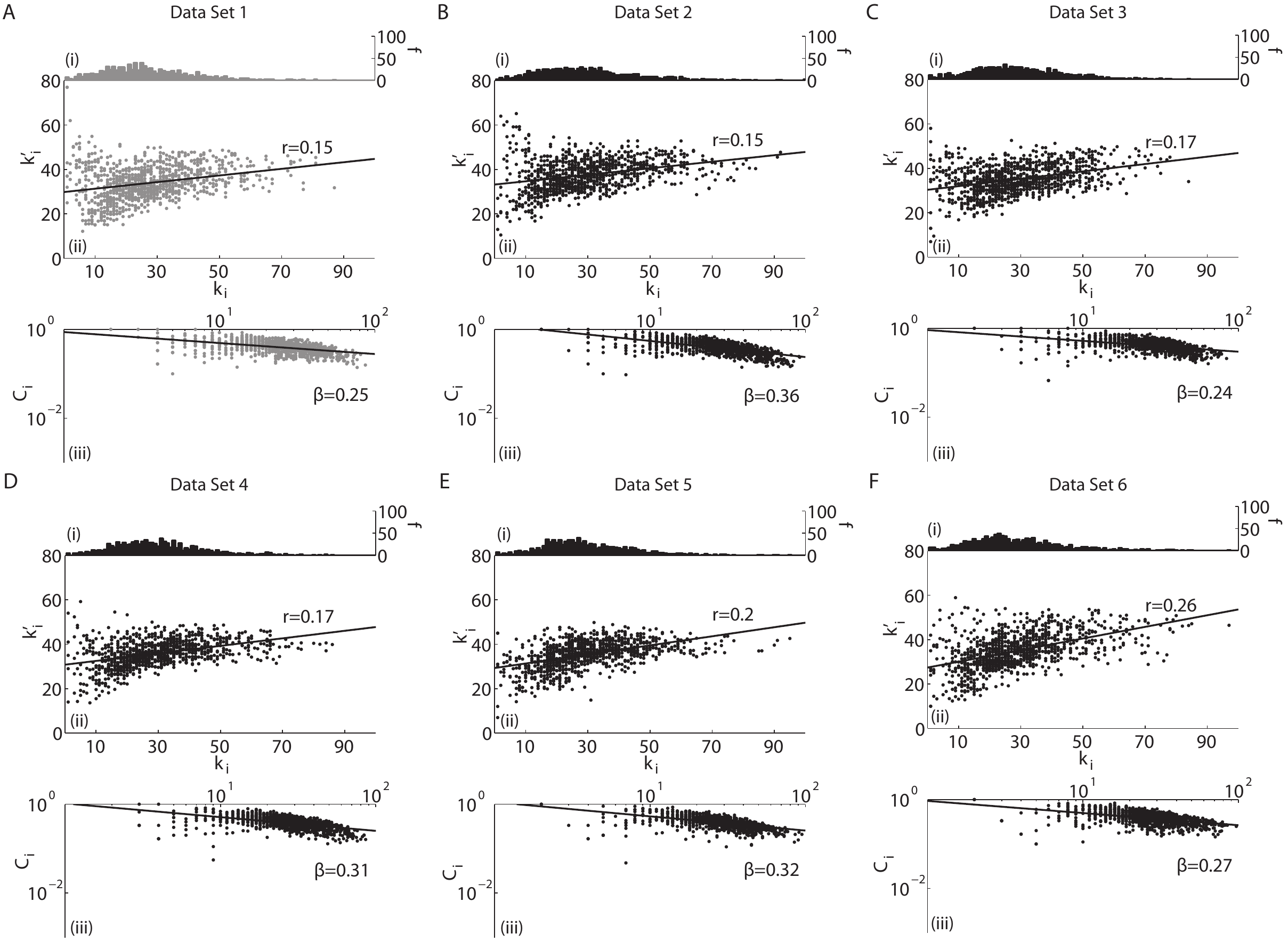}
\caption{\textbf{Reliability of Relational Properties Across Data Sets.} The \emph{(i)} degree distribution (number $f$ of nodes with a given degree $k_{i}$), \emph{(ii)} assortativity (correlation between a node's degree $k_{i}$ and the mean degree of that node's neighbors $k'_{i}$, summarized by parameter $r$), and \emph{(iii)} hierarchy (the relationship between the clustering coefficient $C_{i}$ and the degree $k_{i}$ over all nodes in the network, summarized by parameter $\beta$) for each of the six data sets separately  shown in panels \emph{(A)-(F)}. In panel \emph{(A)}, data set 1 shown in grey was used in the visualizations provided in the main manuscript.}
\end{figure*}

\begin{figure*}[b!]
\centering
\includegraphics[width=0.7\textwidth]{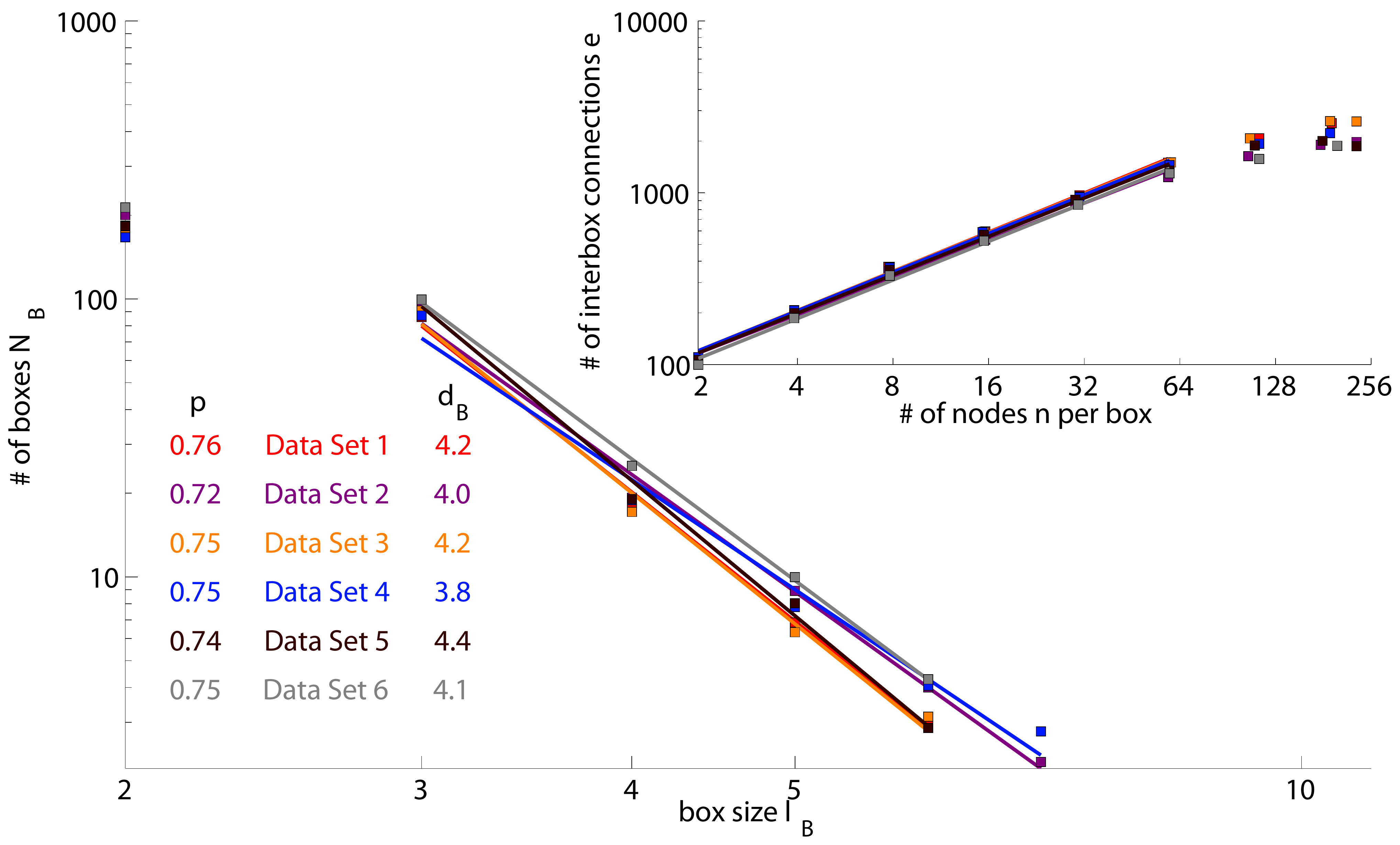}
\caption{\textbf{Reliability of the Topological Dimension Estimates Across Data Sets.} \emph{(Main Panel)} The number of boxes as a function of the topological size of the box, estimated using the box-counting method \cite{Concas2006} (see Methods) for the six empirical brain data sets. \emph{(Inset)} The topological Rentian scaling relationship between the number of edges crossing the boundary of a topological box and the number of nodes inside of the box (see Methods) for the six empirical brain data sets.}
\end{figure*}

\clearpage
\newpage

\bibliographystyle{plos2009}
\bibliography{bibfile2}

\end{document}